%% file: main.tex
\documentclass[sigconf, nonacm]{acmart}
\usepackage{subcaption}
\usepackage{cancel}
\usepackage{multirow}
\usepackage{stfloats}
\usepackage{acmart-taps}
\usepackage{color-edits}
\usepackage{booktabs}
\usepackage{tabularx}
\usepackage{array}
\usepackage{longtable}
\usepackage{comment}

\addauthor{cl}{orange}

{\color{black}}        
{}                    

{\color{blue}}        
{}                    

{\end{sout}}  

\AtBeginDocument{%
  }

\begin{document}

\title[Human--Robot Joint Action]{Understanding Human--Robot Collaboration through Joint Action}

\author{Chengwen Zhang}
\affiliation{%
  \institution{Department of Computer Science and Technology, Tsinghua University}
  \city{Beijing}
  \country{China}}
\email{zcw25@mails.tsinghua.edu.cn}

\author{Jianyang Wang}
\affiliation{%
  \institution{Shanghai Jiao Tong University}
  \city{Shanghai}
  \country{China}}

\author{Chang Liu}
\affiliation{%
  \institution{Department of Computer Science and Technology, University of Cambridge}
  \city{Cambridge}
  \country{United Kingdom}}

\author{Yukun Wang}
\affiliation{%
  \institution{Department of Computer Science and Technology, Tsinghua University}
  \city{Beijing}
  \country{China}}

\author{Yuan Xu}
\affiliation{%
  \institution{The Hong Kong University of Science and Technology (Guangzhou)}
  \city{Guangzhou}
  \country{China}}

\author{Zhoutong Ye}
\affiliation{%
  \institution{Department of Computer Science and Technology, Tsinghua University}
  \city{Beijing}
  \country{China}}

\author{Jie Cai}
\affiliation{%
  \institution{College of Software, Nankai University}
  \city{Tianjin}
  \country{China}}

\author{Chun Yu}
\affiliation{%
  \institution{Department of Computer Science and Technology, Tsinghua University}
  \city{Beijing}
  \country{China}}

\author{Yuanchun Shi}
\affiliation{%
  \institution{Department of Computer Science and Technology, Tsinghua University}
  \city{Beijing}
  \country{China}}

\renewcommand{\shortauthors}{Zhang et al.}
\authorsaddresses{}

\begin{abstract}
  \input{sections/0_abstract.tex}
\end{abstract}

\ccsdesc[500]{Computer systems organization~Robotics}
\ccsdesc[500]{Human-centered computing~Human computer interaction (HCI)}
\keywords{human-robot interaction, systematic review}




\begin{teaserfigure}
  \includegraphics[width=\textwidth]{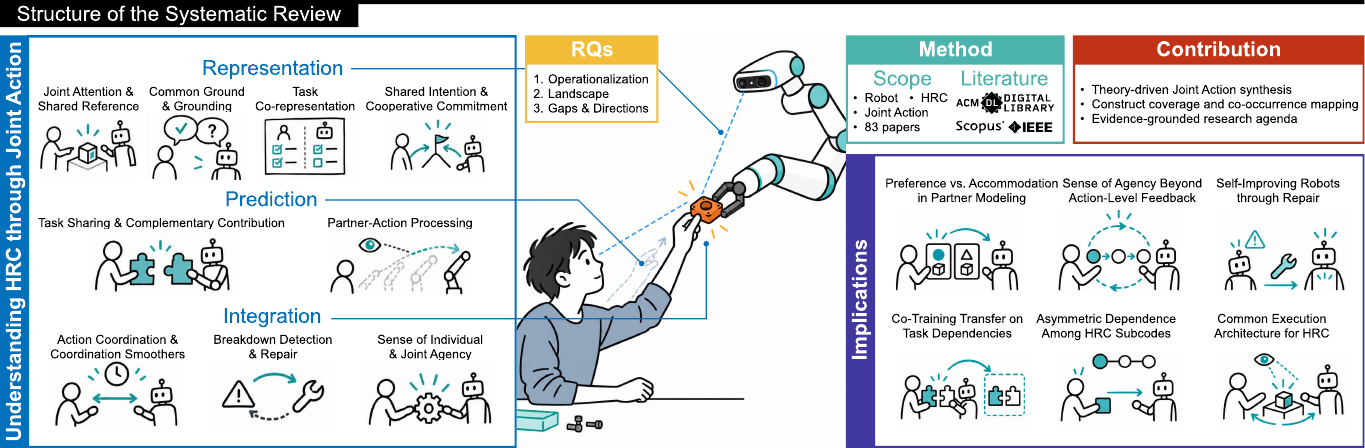}
  \caption{\textbf{Understanding human--robot collaboration (HRC) through Joint Action.} Overview of this systematic review: the Joint Action framework (Representation, Prediction, and Integration) organizing nine constructs, the research questions (RQs) guiding the review, the review method and scope, the resulting contributions, and six implications for future HRC research.}
  \label{fig:teaser}
\end{teaserfigure}

\maketitle
\pagestyle{plain}



\input{sections/1_intro}

\input{sections/2_related}

\input{sections/3_methods}

\input{sections/4_findings}

\input{sections/5_discussion}

\input{sections/6_theory}

\input{sections/7_future}

\input{sections/8_limitations}

\input{sections/9_conclusion}

\begin{acks}
This work was supported by the National Key R\&D Program of China under Grant No. 2024YFB4505500 \& 2024YFB4505501, Beijing Key Lab of Networked Multimedia, Institute for Artificial Intelligence, Tsinghua University (THUAI).
\end{acks}

\bibliographystyle{ACM-Reference-Format}
\bibliography{sample-base}

\appendix

\input{sections/z_appendix}

\end{document}

%% file: sections/0_abstract.tex
Human--robot collaboration requires partners to coordinate interdependent contributions toward a shared goal.
Joint Action theory provides concepts for understanding this coordination, but how these concepts are instantiated and combined across empirical research remains fragmented.
We conducted a systematic review following PRISMA 2020, identifying 83 empirical papers published between 2021 and 2026.
Through theory-driven framework analysis, we organized the literature into nine constructs across three capacities: Representation, Prediction, and Integration.
We analyze how these constructs are distributed and co-occur across studies, revealing where empirical evidence concentrates and where it is absent.
By mapping empirical HRC research onto a unified Joint Action framework, this review offers a coherent overview of HRC studies, identifies gaps in current evidence, and provides guidelines for future research.

%% file: sections/1_intro.tex
\section{Introduction}
\label{sec:intro}

Human--robot collaboration (HRC) brings humans and robots together to accomplish tasks through interdependent contributions~\cite{hoffman2004robots,bauer2008human}. In activities such as handing over an object, transporting a load, or creating an artifact together, each partner's actions shape what the other can do next. Partners need to establish what they are acting on, anticipate one another's contributions, and coordinate when and how to act. When expectations fail, they must also restore the conditions for continuing together. Understanding HRC therefore requires examining the processes that connect individual actions into a shared activity.

Joint Action theory provides a basis for examining these processes. Research on human joint action identifies three functionally related capacities: sharing representations, predicting actions, and integrating the predicted effects of one's own and others' actions~\cite{sebanz2006}. These capacities direct attention to how partners establish a common basis for action, form expectations about one another, and organize complementary contributions in space and time~\cite{vesper2010}. Related accounts of grounding explain how partners establish sufficient evidence of understanding and repair communication when that understanding breaks down~\cite{ClarkBrennan1991}. Together, these perspectives offer concepts for examining how collaboration is established, maintained, and restored across different tasks and robot systems.

Applying these concepts to HRC requires attention to what the available evidence actually establishes. Recognizing a partner's intention does not by itself establish shared intention, and providing information does not establish that the information has been understood. Similarly, coordinated task performance leaves open whether partners hold compatible task representations and how the human experiences control over the joint outcome. Prior work has connected Joint Action theory to interactive robot design, coordination, and the psychological and philosophical challenges of acting with robots~\cite{Curioni2018,IqbalRiek2019,Belhassein2022}. Building on this work requires distinguishing the collaborative mechanisms implemented in a system from the stronger claims about shared understanding, commitment, or agency that its behavior may invite.

Existing reviews provide complementary perspectives on HRC, including industrial safety and interfaces~\cite{Villani2018}, robot learning~\cite{Mukherjee2022}, and the broader organization of collaboration, communication, and cognition~\cite{Singh2026}. More focused reviews examine communication, intention-based systems, and human autonomy and agency~\cite{Salehzadeh2022,ZhangDoyle2023,Glawe2026}, while Joint Action reviews connect theoretical mechanisms to HRI and compare interaction with human and robot partners~\cite{Belhassein2022,Navare2026}. These contributions establish a foundation for a further question: how do empirical HRC studies instantiate distinct Joint Action constructs, and which constructs are studied together? Addressing this question requires bringing together studies that use different terminology while preserving distinctions among mechanisms that may appear similar in system descriptions. It also requires including relevant studies that investigate collaborative mechanisms without explicitly adopting the term \emph{joint action}.

To address this question, we conducted a systematic review following PRISMA 2020~\cite{page2021prisma}. Our review includes 83 empirical papers published between 2021 and August 10, 2026, in which humans and robots contribute to a shared task through interdependent or coordinated actions and collaborative mechanisms are a central focus. We combined theory-driven coding with open coding of how each construct was operationalized. The resulting framework comprises nine constructs organized across \emph{Representation}, \emph{Prediction}, and \emph{Integration}, with 29 subcodes describing their implementation in the reviewed studies. These capacities provide complementary perspectives on joint activity and can operate concurrently. Our investigation addresses three research questions:

\begin{itemize}
    \item \textbf{RQ1:} How are Joint Action constructs instantiated in empirical human--robot collaboration research?
    \item \textbf{RQ2:} How is research attention distributed across these constructs, and how do their operationalizations co-occur within studies?
    \item \textbf{RQ3:} What evidence gaps and research directions emerge from examining human--robot collaboration through Joint Action?
\end{itemize}

Our synthesis shows that the reviewed literature concentrates on a common execution architecture of partner-action processing, task sharing, and action coordination, while shared intention and cooperative commitment are less frequently addressed. Co-occurrence analysis reveals asymmetric dependencies among subcodes: task division, monitoring, and action prediction function as structural prerequisites that most other collaborative mechanisms presuppose. Human adaptation to robot limitations emerges as a recurring enabler of smooth execution, yet whether it reflects genuine preference or temporary accommodation remains largely unexamined. These findings motivate closer examination of whether successful execution is accompanied by compatible task understanding, experienced agency, and manageable demands on the human partner. We use this synthesis to identify directions for partner modeling, agency evaluation, learning through repair, transfer in co-training, and the management of pauses, handoffs, and withdrawal. This paper makes three contributions:

\begin{itemize}
    \item A theory-driven synthesis of 83 empirical HRC papers that distinguishes nine Joint Action constructs and characterizes their operationalizations across Representation, Prediction, and Integration.
    \item An analysis of construct coverage and subcode co-occurrence that maps the distribution and combinations of collaborative mechanisms investigated in the reviewed literature.
    \item An evidence-grounded research agenda that connects gaps in current implementations and evaluations to questions about human understanding, agency, adaptation, and the sustained organization of joint activity.
\end{itemize}

%% file: sections/2_related.tex
\section{Related Work and Background}
\label{sec:related}

\subsection{Scope}
\label{sec:scope}
This section defines the key terms used in this paper to establish a common ground for our discussion and clarify the scope of this study.

\textbf{Robots} are conventionally characterized as systems that perceive their environment and act toward task goals, with autonomy treated as a defining property of robotic systems \cite{goodrich2008human,iso8373}. ISO 8373:2021 defines a robot as a programmed actuated mechanism with a degree of autonomy for locomotion, manipulation, or positioning \cite{iso8373}. HRI research includes physically co-present and virtually represented robots \cite{li2015benefit} and Wizard-of-Oz (WoZ) control, which is used to study interaction when robot behavior is not fully autonomous \cite{riek2012wizard}. We define a robot as an artificial agent that is presented as a robotic entity and participates in a joint task through task-relevant actions that affect the progression of the collaboration. The definition covers physically embodied robots, virtually represented robots, and WoZ implementations. Tool-like controllers and interface-only systems remain classified as human instruments \cite{scholtz2003theory,bradshaw2009tools}.


\textbf{Human--robot collaboration} refers to settings in which humans and robots work together toward a shared task goal \cite{hoffman2004robots,bauer2008human,Ajoudani2018}. HRC emphasizes the robot's participation in task execution and the contributions of both human and robot to accomplishing the task \cite{bauer2008human}. Robots in HRC are described as partners or teammates whose actions form part of the joint activity \cite{hoffman2004robots,bradshaw2009tools}. HRI taxonomies distinguish collaborative roles from supervision or operation by a shared overall goal and contributions from both parties \cite{onnasch2021taxonomy}. Contributions may be unequal, and either party may be unable to complete the task alone. This survey defines HRC as a joint task involving at least one human and one robot, where both participate in task execution and contribute toward a shared goal.

\subsection{Joint Action Theory}
\label{sec:joint-action-theory}

Joint action is commonly defined as social interaction in which two or more individuals coordinate their actions in space and time to bring about a change in the environment \cite{sebanz2006}. Sebanz, Bekkering, and Knoblich proposed that successful joint action depends on three functionally related capacities: sharing representations, predicting actions, and integrating the predicted effects of one's own and others' actions \cite{sebanz2006}. Their review connects these capacities to joint attention, action observation, task sharing, action coordination, and agency. This survey uses these three capacities as the organizing structure for the theory-driven synthesis, labeled \emph{Representation}, \emph{Prediction}, and \emph{Integration}. The three capacities can operate in parallel.

\textbf{Representation} concerns the shared representations that allow partners to participate in the same joint activity. At the perceptual level, joint attention and shared reference allow partners to orient toward the same object or event; Sebanz et al. describe this as a form of perceptual common ground \cite{sebanz2006}. At the task level, joint-action research shows that actors may represent not only their own task but also the task assigned to a partner, supporting task co-representation \cite{sebanz2006,vesper2010}. Clark and Brennan extend this representational basis to shared understanding: through grounding, partners establish and update evidence that their understanding is sufficient for the current purpose \cite{ClarkBrennan1991,Klein2005}. Bratman's planning account further emphasizes interlocking intentions, meshing subplans, and sustained commitment and mutual support in joint activity \cite{bratman1992,bratman2014}. These accounts help distinguish participation in a common task from the formation of shared representations. Accordingly, our identification of shared intention and common ground relies on evidence of shared planning or the establishment and updating of mutual understanding.

\textbf{Prediction} concerns how partners anticipate one another's actions. Sebanz et al. identify two complementary sources of such prediction. Action observation provides online information about what another actor is doing and is likely to do next, whereas knowledge of the partner's task allows expectations about what the partner should do to be formed before the relevant action is observed \cite{sebanz2006}. Vesper et al. also place monitoring and prediction at the core of their minimal architecture for joint action \cite{vesper2010}. In our framework, \emph{Partner-Action Processing} covers perceiving, monitoring, interpreting, and predicting the partner's unfolding actions, while \emph{Task Sharing and Complementary Contribution} describes the roles and contributions from which task-based expectations about partner behavior are derived. This distinction allows us to separate expectations derived from task structure from online judgments based on the partner's unfolding behavior, and thus to examine where predictions about partner action originate.

\textbf{Integration} concerns how partners organize their actions into a joint activity that can accomplish the task. Sebanz et al. characterize action coordination as using shared representations and action prediction to select and time complementary actions \cite{sebanz2006}. Vesper et al. further describe coordination smoothers that reduce the demands of online coordination \cite{vesper2010}. Clark and Brennan show how breakdowns in shared understanding or coordination can be detected and repaired, thereby restoring the joint activity \cite{ClarkBrennan1991}. Tightly coupled joint actions also raise the question of whether their effects are attributed to oneself or to the acting dyad or group; work on agency distinguishes self-agency from collective experiences of agency \cite{Pacherie2012,Loehr2022}. These accounts direct our analysis toward the relations formed between partners' actions during execution. Action coordination therefore concerns how their actions are organized in relation to one another, repair concerns how joint activity is restored after a breakdown, and agency concerns how the outcomes of joint action are experienced and attributed.


\subsection{Existing Reviews and the Scope of This Survey}
\label{sec:existing-reviews}

Existing reviews of human--robot interaction and collaboration often organize the literature around engineering or system-oriented perspectives. Reviews of industrial human--robot collaboration mainly discuss safety, interaction, and application issues \cite{Villani2018,Ajoudani2018}. Reviews of robot learning for collaboration, in contrast, primarily organize relevant methods around how robots learn, adapt, and generalize collaborative behaviors \cite{Mukherjee2022}.

Singh and colleagues organize human--robot teaming around collaboration, communication, and cognition and propose a 3Cs rating system \cite{Singh2026}. This framework groups several constructs that can be further distinguished in Joint Action theory into broader categories. For example, shared understanding includes shared reference, common ground, and task co-representation. In this survey, these constructs are coded separately.

Other reviews focus on a particular interaction capability, such as purposeful communication \cite{Salehzadeh2022}, intention-based systems \cite{ZhangDoyle2023}, and human autonomy and sense of agency \cite{Glawe2026}. These single-construct reviews can provide in-depth discussion of specific issues, but they make it more difficult to show how multiple Joint Action constructs are instantiated and combined within the same HRC system. At the same time, some phenomena that appear closely related in HRI have distinct meanings in Joint Action theory, such as intent recognition and shared intention, as well as information provision and grounding. We preserve these theoretical distinctions in our analysis.

A smaller set of reviews uses Joint Action as a primary lens for analyzing HRC. Curioni, Knoblich, and Sebanz introduced mechanisms from human Joint Action research into HRI and discussed their implications for interactive robot design \cite{Curioni2018}. Iqbal and Riek examined coordination issues in human--robot teaming from the perspectives of Joint Action and dynamical systems \cite{IqbalRiek2019}. Belhassein and colleagues analyzed human--robot joint action from psychological and philosophical perspectives and emphasized the role of communication \cite{Belhassein2022}. Navare and colleagues systematically compared Joint Action with human and robot partners \cite{Navare2026}. These reviews have established the connection between Joint Action and HRI. This survey further examines how HRC studies instantiate and combine constructs already distinguished in Joint Action theory, including studies that do not explicitly use the term \emph{joint action} but nevertheless engage with the corresponding mechanisms.

%% file: sections/3_methods.tex
\section{Literature Search and Study Selection}

This systematic review follows the Preferred Reporting Items for Systematic Reviews and Meta-Analyses (PRISMA) 2020 guidelines~\cite{page2021prisma}.

\subsection{Search Strategy}
A complete list of all databases, search results, and inclusion counts is provided in the appendix.

\subsubsection{Search Query Formulation.}
After surveying a broad body of relevant literature, we found that human–robot collaboration spans multiple research traditions, including HRI, robotics, cognitive science, and human factors, with no consistent terminology for describing collaborative relationships across these communities. We therefore began with an exploratory search in the ACM Digital Library (ACM DL)\footnote{The search was conducted on August 10, 2026.}, using the broad query robot* AND human* to examine how relevant studies were described. This initial search proved overly inclusive, retrieving a large number of studies concerned with general human–robot interaction, robot control, or user studies, rather than collaboration specifically.

We then drew on the core characteristics of joint action and explored terms such as joint action, shared goal, interdependence, and coordination to narrow the search. However, our initial reading showed that many studies consistent with our definition of human–robot collaboration did not explicitly use terms such as joint action, shared goal, or interdependence. Instead, they more commonly described collaborative behavior using collaboration, cooperation, communication, or coordination~\cite{index5,index36,gilbert2025human,khan2025errors}. We therefore adopted these four terms as the primary set of collaboration-related keywords.

To further refine the query, we examined recurring terms in the titles, abstracts, and keywords of a set of representative papers judged to be relevant to our review. After excluding overly general terms such as robot, human, and task, we observed that predict, anticipate, and intent, together with their morphological variants, appeared frequently~\cite{index62,index74,laplaza2025enhancing,index28}. As these terms appeared to capture aspects of prediction and intent understanding that are often involved in collaborative interaction, we initially incorporated them into the query:

(robot* AND human) AND (collaborat* OR cooperat* OR communicat* OR coordinat*) AND (predict* OR anticipat* OR intent*)

However, this additional constraint substantially reduced the number of retrieved studies. A subsequent check against known relevant papers showed that many studies satisfying our criteria for shared goals and interdependent action instead framed their contributions in terms of task allocation, mutual adaptation, shared control, collaborative manipulation, or team coordination, without explicitly using prediction, anticipation, or intent in their titles, abstracts, or keywords~\cite{index98,perez2024experimental,ramachandruni2024uhtp,fuster2025negotiation}. Treating these terms as mandatory search criteria would therefore have reduced recall and systematically excluded relevant work. Based on these iterations, we adopted the following final query:

\begin{equation}
\begin{aligned}
&(\mathit{robot*} \ \mathbf{AND}\ \mathit{human}) \\
&\mathbf{AND}\ (\mathit{collaborat*} \ \mathbf{OR}\ \mathit{cooperat*} \ \mathbf{OR}\ 
\mathit{communicat*} \ \mathbf{OR}\ \mathit{coordinat*})
\end{aligned}
\end{equation}

The first term set ensures that retrieved studies involve both humans and robots, while the second captures a broad range of terminology used to describe human–robot collaboration. We deliberately did not require theoretical terms such as joint action, shared goal, or interdependence, because studies may instantiate these properties without explicitly naming them. Instead, these conceptual criteria were assessed during the subsequent manual screening stage.

\subsubsection{Database Selection}
Given the interdisciplinary nature of human–robot collaboration research, which spans HCI, robotics, psychology, and behavioral science, we selected the ACM Digital Library (ACM DL) as our primary database to cover core research in HCI and HRI. To mitigate potential disciplinary bias arising from reliance on a single database, and following consultation with domain experts, we supplemented our search with RO-MAN, a major conference on human–robot interaction, as well as four representative journals: Nature, Science Robotics, Computers in Human Behavior, and the International Journal of Social Robotics. RO-MAN and Science Robotics broadened our coverage of robotics research; the International Journal of Social Robotics provided work on social robots and social aspects of human–robot interaction; Computers in Human Behavior contributed perspectives from psychology, behavioral science, and human factors; and Nature was included to capture relevant work published in high-impact multidisciplinary venues.

\subsection{Screening and Selection}
\subsubsection{Inclusion and Exclusion Criteria.}
\label{sec:inclusion}
The literature screening process was guided by predefined inclusion and exclusion criteria to ensure that the final set of studies aligned with the context of human--robot collaboration examined in this review. Specifically, papers were required to have been published between 2021 and 2026, written in English, peer-reviewed, and available in full text, and to include at least one empirical study in which the human and robot shared a common task goal and their actions were interdependent or coordinated.

Consequently, studies that did not meet these criteria were excluded. We excluded publications that did not satisfy the requirements regarding publication period, language, full-text availability, or publication type, such as review articles and short papers. Studies involving only humans or only robots were excluded \cite{hirano2026ball}. For studies involving both humans and robots, we also excluded those in which the two did not share a common task goal \cite{onishi2026being}, or in which the interaction primarily consisted of one-way control, demonstration, or information transfer, without substantive action interdependence or bidirectional coordination~\cite{maccio2024kinesthetic}.

In addition, some papers used human–robot collaboration merely as a task or experimental context, while their primary research questions focused on topics such as trust and individual capability assessment, task allocation and scheduling, control methods, or interface design, rather than on the mechanisms of human–robot collaboration itself \cite{hald2021human}. Therefore, these papers were also excluded.

\subsubsection{Resolution of Disagreements.}
During the screening process, disagreements between the two primary reviewers were resolved through a staged procedure. The two reviewers first discussed each discrepant decision and attempted to reach consensus based on the predefined inclusion and exclusion criteria. If consensus could not be reached, the case was referred to a third reviewer for further review and adjudication.

Overall, 18 papers resulted in disagreements regarding inclusion or exclusion. During full-text screening, the two reviewers made the same decision for 177 of the 195 papers, corresponding to an overall agreement rate of 90.77\% and a Cohen’s $\kappa$ of 0.811, indicating a high level of inter-reviewer agreement. Disagreements mainly concerned several boundary cases: (1) whether the human played a substantive role in the joint task; (2) whether studies primarily designed to validate algorithms made a substantive contribution to understanding human–robot collaboration; (3) whether experiments conducted in virtual reality, simulation environments, or through video-based stimuli could provide sufficient evidence of human–robot collaboration; and (4) whether the robot embodiment and interaction medium fell within the scope of robots defined in this review.

For example, Khan et al. \cite{khan2025errors} investigated how errors made by a virtual collaborative robot during gearbox assembly affected interaction fluency and workload. However, participants were mainly asked to detect robot errors and choose between automatic recovery and training modes, or to indicate the correct action sequence. They did not share responsibility with the robot for carrying out the assembly task itself, and the paper was therefore excluded. Naik et al. \cite{naik2021multi} primarily examined how a robot proactively approached a person and delivered a drink. The human participant served mainly as a target to be detected and approached, rather than contributing actions to the planning or completion of a shared task, and the paper was likewise excluded. Although Lima et al. \cite{lima2023assistive} and Shah et al. \cite{shah2023robot} respectively involved human–robot co-manipulation and robotic assistance under cognitive fatigue, their main contributions concerned an assistive force-control architecture and a reinforcement-learning-based adaptive control strategy. In both cases, the experiments were primarily used to evaluate control performance and its effects on task-level outcomes rather than to investigate mechanisms of human–robot collaboration. These papers were therefore excluded as well. Roy et al. \cite{roy2023need} and Moon et al. \cite{moon2021design} mainly studied situational awareness and hesitation behaviors by asking participants to watch videos of robot behavior or evaluate prerecorded robot motions. Because participants did not engage in a real-time joint task or reciprocal interaction with the robot, these studies did not provide the form of collaboration evidence required by our criteria.

Additional disagreements concerned the boundary between robotic and non-robotic agents. In Roche and Saint-Bauzel \cite{roche2022study}, the “partner” was a virtual algorithmic agent constructed from human behavioral data. Chiou et al. \cite{chiou2022towards} studied collaboration with a purported remote robot in a virtual search-and-rescue environment through text chat, maps, and video. In both cases, there was no robot system capable of performing observable actions as part of the joint task, and the papers were excluded. In contrast, although the systems studied by Pietras et al. \cite{index27}, Angelopoulos et al. \cite{index51}, and Guinot et al.~\cite{index122} did not involve conventional physical robots, or were evaluated in virtual environments, the robotic agents themselves performed actions that contributed directly to the shared task and engaged in reciprocal coordination with human participants. These papers were therefore included.

After final adjudication, 9 of the 18 disputed papers were included and 9 were excluded.

\begin{figure}[h]
    \centering
    \includegraphics[width=\columnwidth]{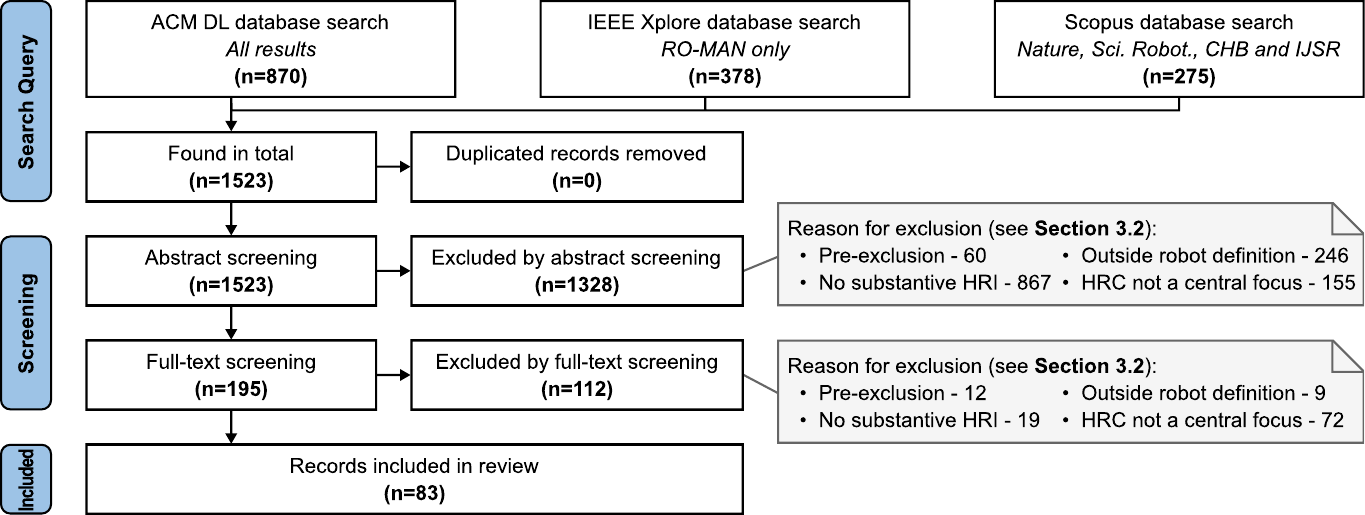}
    \caption{PRISMA flow diagram outlining the literature screening and inclusion process for this systematic review.}
    \label{fig:PRISMA}
\end{figure}

\begin{figure}[h]
    \centering
    \includegraphics[width=\columnwidth]{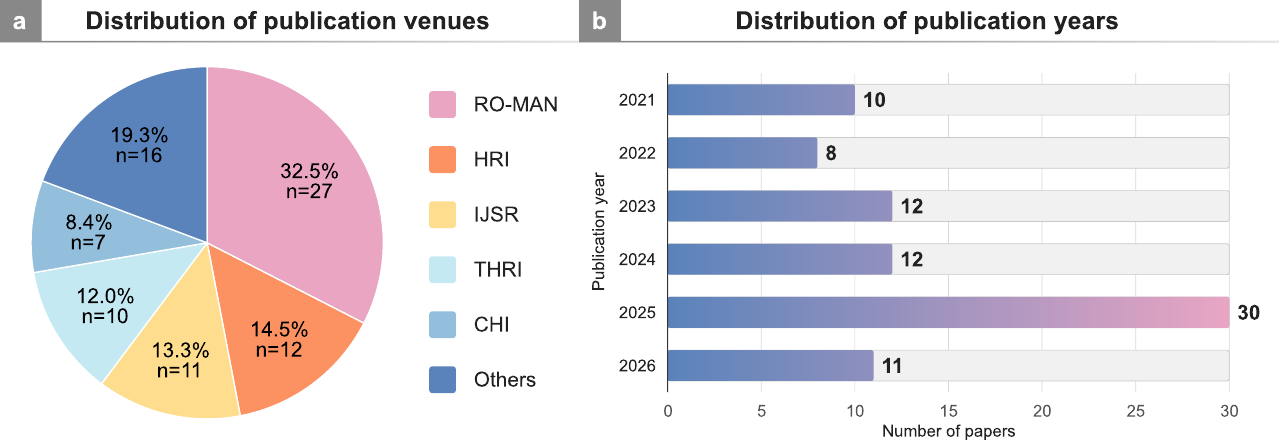}
    \caption{Overview of publication venues and years. (a) Distribution of included papers by publication venue. (b) Annual number of included papers. The 2026 data include publications available up to August 10, 2026.}
    \label{fig:Overview}
\end{figure}

\subsection{Final Corpus}
\subsubsection{Paper Selection Process.}
A complete PRISMA flow diagram illustrating the paper selection process is provided in Figure~\ref{fig:PRISMA}. The initial database search yielded 1,523 unique records. During title and abstract screening, 1,328 irrelevant records were excluded, leaving 195 papers for full-text eligibility assessment. In the final full-text screening stage, all 195 papers were read in their entirety, and 112 were excluded after further evaluation. The main reasons for exclusion were pre-exclusion criteria (12 studies), not meeting our definition of a robot (9 studies), the absence of substantive human–robot interaction (19 studies), and human–robot collaboration mechanisms not being a central focus of the study (72 studies). This multi-stage selection process resulted in a final corpus of 83 studies.

\subsubsection{Overview of Included Papers.}
The distribution of publication venues and years for the papers included in this review is shown in Figure~\ref{fig:Overview}. The 83 papers were published across major conferences and journals in human–computer interaction, robotics, and social robotics. The IEEE International Conference on Robot and Human Interactive Communication (RO-MAN) contributed the largest number of papers, followed by the ACM/IEEE International Conference on Human-Robot Interaction (HRI), the International Journal of Social Robotics, ACM Transactions on Human-Robot Interaction (THRI), and the ACM CHI Conference on Human Factors in Computing Systems (CHI). Together, these five venues accounted for 76 papers, representing 91.6\% of the included literature. Annual counts were 10, 8, 12, and 12 papers from 2021 to 2024, 30 in 2025, and 11 in 2026 (up to August 10).

\subsection{Theory-Driven Coding and Analysis}
\label{sec:coding-analysis}

After screening yielded a final corpus of 83 papers, we conducted a theory-driven framework analysis to examine how HRC research operationalizes constructs from established Joint Action theory. Our analysis combined directed qualitative content analysis with a best-fit framework approach: candidate constructs were first specified a priori from established theory and then refined through pilot coding in terms of their operational definitions, organization, and discriminant boundaries \cite{HsiehShannon2005,Carroll2013,Gale2013}.

\subsubsection{Pilot Sampling}
\label{sec:pilot-sampling}

We constructed the pilot sample using maximum-variation sampling \cite{Suri2011,Benoot2016}. A sampling matrix was developed from directly observable characteristics of the 83 included papers, from which we selected 24 development papers and 12 held-out papers. The development set was coded in three rounds of eight papers, respectively used to establish basic coding criteria, examine boundaries between adjacent constructs, and test the framework on more complex cases. The 12 held-out papers were selected and locked before coding began and did not participate in codebook revision. They were used to assess whether the resulting coding rules could be applied consistently to papers that had not contributed to framework development. The sampling dimensions and stage-specific design are detailed in Tables~\ref{tab:app_sampling_matrix} and~\ref{tab:app_sampling_design}; the development and held-out samples are listed in Tables~\ref{tab:diagnostic_papers} and~\ref{tab:app_heldout_sample}.

\subsubsection{Iterative Codebook Development}
\label{sec:codebook-development}

In Round 0, we extracted 11 initial candidate constructs from a separate corpus of established Joint Action theory. Pilot coding was used to refine the organization and coding boundaries of these theory-derived constructs. When a potential gap was identified, the theory corpus was re-examined, and structural changes were made only when supported by established theory.

Round 1 examined the identifiability and organization of the candidate constructs. Action Observation, Action Prediction, and Action Monitoring were integrated into Partner-Action Processing because they all concerned the online processing of a partner's actions. This round also clarified the distinctions between Task Co-representation, Task Sharing, and Action Coordination.

Round 2 focused on the boundaries between adjacent constructs using cases prone to false-positive coding. For example, gaze was coded as Joint Attention only when it established or confirmed a shared referent supporting subsequent joint action. The boundary cases and final decision rules are detailed in Tables~\ref{tab:app_boundary_pressure} and~\ref{tab:app_boundary_rules}. This round also showed that Joint Action Goal (JA1) could not be consistently distinguished from the shared task goal already present in the study setting and was therefore not retained as an independent primary code.

Round 3 examined whether the revised framework still contained coverage gaps in more complex cases. Failure-and-recovery cases revealed that the existing constructs did not adequately capture the restoration of joint activity after a breakdown. We therefore re-examined the Grounding and Joint Activity literature and confirmed that Repair is an established mechanism in this theoretical tradition. Breakdown Detection and Repair was consequently added to the framework \cite{ClarkBrennan1991,Klein2005}. This round produced the release-candidate codebook.

\subsubsection{Held-out Test and Final Framework}
\label{sec:final-framework}

Finally, two coders independently applied the release-candidate manual to the 12 held-out papers to assess whether the construct definitions and coding boundaries remained stable on cases that had not participated in codebook development. At this stage, only wording clarifications that did not alter the construct structure were permitted. After the necessary theoretical checks were completed, the codebook was frozen and applied to the full corpus. The final framework comprised nine multi-label primary codes, organized into three levels based on joint action theory: Representation, Prediction, and Integration. Table~\ref{tab:final-framework} summarizes this three-level structure and the operational focus of each code. Detailed definitions and inclusion/exclusion criteria are provided in Table~\ref{tab:app_detailed_codebook}, with discriminant boundary rules provided in Table~\ref{tab:app_boundary_rules}.

\begin{table*}[h]
\centering
\caption{The final theory-derived coding framework, comprising nine multi-label primary codes organized across three levels of joint action.}
\label{tab:final-framework}
\footnotesize
\renewcommand{\arraystretch}{1.15}

\begin{tabularx}{\textwidth}{
    >{\raggedright\arraybackslash}p{2.3cm}
    >{\raggedright\arraybackslash}p{5.1cm}
    >{\raggedright\arraybackslash}X
}
\toprule
\textbf{Level} & \textbf{Primary Code} & \textbf{Operational Focus} \\
\midrule

\multirow{4}{*}{\textbf{Representation}}
& JA4 Joint Attention and Shared Reference
& Coordinating attention to establish a shared referent that supports subsequent joint action. \\

& JA5 Common Ground and Grounding
& Establishing and updating mutually usable task-relevant knowledge through evidence of understanding. \\

& JA3 Task Co-representation
& Representing one's own task, the partner's task, joint-task rules, and relations among contributions. \\

& JA2 Shared Intention and Cooperative Commitment
& Supporting joint planning and execution through shared intentional structure and continued cooperative commitment. \\

\midrule

\multirow{2}{*}{\textbf{Prediction}}
& JA7 Task Sharing and Complementary Contribution
& Dividing a joint task into interdependent roles or contributions that complement one another. \\

& JA6 Partner-Action Processing
& Perceiving, predicting, and monitoring partner actions or joint outcomes online to support coordination. \\

\midrule

\multirow{3}{*}{\textbf{Integration}}
& JA8 Action Coordination and Coordination Smoothers
& Coordinating actions during execution, including mechanisms that reduce coordination difficulty. \\

& JA9 Breakdown Detection and Repair
& Detecting breakdowns in joint activity and restoring coordination through repair. \\

& JA10 Sense of Individual and Joint Agency
& Experiencing or attributing control, authorship, and ownership concerning one's contribution and the joint outcome. \\

\bottomrule
\end{tabularx}
\end{table*}

\subsubsection{Full-Corpus Coding and Reliability}
\label{sec:full-corpus-coding}

After the final coding framework was established, we applied it to all 83 included papers at two interrelated levels. At the evidence level, coders identified full-text passages supporting a Joint Action construct and recorded the original excerpt, page number, the authors' terminology, and the corresponding primary code. At the paper level, these evidence units were aggregated to determine which constructs were addressed in each paper. Primary-code assignment followed the frozen coding manual throughout and was not modified after the framework was finalized. For evidence already assigned to a primary code, we further conducted open coding to characterize how each construct was operationalized in HRC research, iteratively comparing and consolidating semantically similar subcodes. The two coders, with backgrounds in computer science and industrial design respectively, studied the Joint Action theory corpus and the coding manual before independently coding all 83 papers in full text. Inter-rater reliability was calculated from the two independent results before disagreement adjudication; for each dimension, we report raw agreement and Cohen's $\kappa$ in Table~\ref{tab:ja_irr}.

JA2 (Shared Intention and Cooperative Commitment) and JA8 (Action Coordination and Coordination Smoothers) showed relatively high disagreement. JA2 was difficult because shared intention was often implicit and had to be inferred from the interaction. For instance, Yang and Mavrogiannis's study was coded as JA2 because the human and robot gradually established a shared strategy for passing an obstacle during collaborative transport \cite{index40}; by contrast, Al-Saadi et al., though also studying collaborative transport with post-conflict action adjustment, provided evidence mainly concerning failure recovery, insufficient to support shared intention \cite{index98}. JA8 disagreements concerned the boundary between general interaction behavior and action coordination: Kimoto et al.'s Help-Estimator was coded as JA8 because the robot's pause after requesting help and the human's task switch upon acceptance created a clear action linkage \cite{index114}; whereas Andriella et al. adapted information to the user's state but did not organize the execution of both agents' actions, and was therefore excluded \cite{index31}. In summary, JA2 required sufficient evidence that partners were forming or maintaining a shared plan, while JA8 required a clear relation of mutual constraint between their actions during execution. Papers with inconsistent coding were discussed by the two coders; unresolved cases were adjudicated by a third researcher to produce the final corpus coding.

\begin{table*}[h]
\centering
\caption{Inter-rater reliability of paper-level Joint Action coding.}
\label{tab:ja_irr}
\renewcommand{\arraystretch}{1.18}

\begin{tabularx}{\textwidth}{
    >{\raggedright\arraybackslash}X
    >{\centering\arraybackslash}p{3.0cm}
    >{\centering\arraybackslash}p{2.8cm}
}
\toprule
\textbf{Joint Action Dimension}
& \textbf{Raw Agreement}
& \textbf{Cohen's $\kappa$} \\
\midrule

JA2 Shared Intention and Cooperative Commitment
& 77.1\%
& 0.146 \\

JA3 Task Co-representation
& 80.7\%
& 0.614 \\

JA4 Joint Attention and Shared Reference
& 81.9\%
& 0.618 \\

JA5 Common Ground and Grounding
& 86.7\%
& 0.666 \\

JA6 Partner-Action Processing
& 95.2\%
& 0.644 \\

JA7 Task Sharing and Complementary Contribution
& 91.6\%
& 0.543 \\

JA8 Action Coordination and Coordination Smoothers
& 84.3\%
& 0.329 \\

JA9 Breakdown Detection and Repair
& 92.8\%
& 0.856 \\

JA10 Sense of Individual and Joint Agency
& 88.0\%
& 0.700 \\

\bottomrule
\end{tabularx}
\end{table*}

%% file: sections/4_findings.tex
\section{Understanding HRC through Joint Action}\label{sec:findings}

We organize the findings into three layers: Representation, Prediction, and Integration. Figure~\ref{fig:ja-category-distribution} summarizes the distribution of coded categories within JA2--JA10, while Figure~\ref{fig:ja-category-cooccurrence} shows how these categories co-occur across the reviewed papers.

\subsection{Representation}\label{sec:representation}
Representation concerns how partners establish the cognitive basis required for acting together. Specifically, humans and robots may form representations of shared objects, each other's knowledge states, their respective task roles, and the joint activity itself.
Joint action theory establishes that partners need representational coordination---shared reference, common ground, compatible task models, and cooperative orientation~\cite{ClarkBrennan1991}. Across the coded HRC literature, these requirements are realized through specific engineering means: shared reference is anchored in the environment (Section~\ref{sec:ja4}), grounding is established through procedures (Section~\ref{sec:ja5}), task structure is encoded in explicit models (Section~\ref{sec:ja3}), and cooperative orientation is sustained through behavioral responsiveness rather than rich shared plans (Section~\ref{sec:ja2}).
The central representational problem in HRC is therefore whether representations are sufficiently compatible to sustain the joint activity.


\subsubsection{Joint Attention and Shared Reference}\label{sec:ja4} 

Joint Attention and Shared Reference is the Representation-level mechanism for aligning on the same object or event. It concerns whether both partners can treat an object, location, event, or action target as the thing they are currently referring to together. HRC recasts this as an engineering problem: how to establish an actionable shared referent. Two features recur. First, existing work typically uses multiple channels: language, bodily cues, task history, and spatial displays that remain in place.
Second, HRC relies heavily on externalization:
human--human joint action often leaves the shared referent to be maintained through internal coordination processes,
while many HRC systems put the reference into the shared environment so that both partners can organize later action around the same external mark.
Shared reference therefore has to be constructed.
We analyze shared reference through three mechanisms that are not strictly mutually exclusive: Language, Deictic Cues, and Spatial Anchoring.

\paragraph{Language.}\label{sec:ja4-language}
Language specifies the object, location, action, or task state under discussion through both direct and context-dependent reference. For direct reference, robots can ask which task module has failed \cite{index1}, request clarification when object descriptions are underspecified \cite{index21}, or use speech and discrete options to specify the next action \cite{index50,index58,index92}. Fixed verbal prompts can also bind instructions to particular objects or actions \cite{index37,index39,index114}. For context-dependent reference, systems interpret ellipsis from prior actions and task history \cite{index26}, resolve spatial language from trajectory context \cite{index5}, or generate referring expressions from the user's beliefs and spatial perspective \cite{index31,index51,index109}. These studies locate the interpretation of a referring expression within the unfolding task: task history and perspective information help resolve what the utterance leaves unspecified \cite{index5,index26,index31}.

\paragraph{Deictic Cues.}\label{sec:ja4-deixis}
Deictic cues locate abstract language or intent on a concrete referent in shared physical space. In robot-to-human deixis, robots can use head and body orientation to indicate interaction targets \cite{index9,index67,index86}, combine gaze and pointing to place verbal instructions in physical space \cite{index83}, or use deictic gestures to mark candidate objects that need confirmation or correction \cite{index147}. In human-to-robot deixis, users can point to designate target objects \cite{index21}, or use gestures to specify handover locations \cite{index104}. These cues connect spoken or task intentions to visible spatial referents, helping partners shift attention and prepare actions faster \cite{index83}.

\paragraph{Spatial Anchoring.}\label{sec:ja4-spatial}
Spatial anchoring externalizes the shared referent into space in two main ways: by pinning the intended object to the environment or an interface, or by reporting its location relative to the human body. External anchoring uses persistent spatial marks to hold shared objects and action targets in place, including shared displays that give the team a common reference frame \cite{index14,index15}, coordinate frames that show planned robot poses \cite{index27}, and LEDs \cite{index37,index39}, AR overlays \cite{index41}, or environmental tags \cite{index71} that bind information to particular locations or objects. Body-relative anchoring uses the human body, or the spatial relation between human and robot, as the coordinate frame. A handover point, for example, can be given as direction and distance from the human hand and conveyed by vibration \cite{index74}; relative position can also be presented through left and right audio channels \cite{index128}. In both cases the shared referent persists beyond language or momentary gaze as an object both partners can continue to access, supporting later attention and action preparation.

\subsubsection{Common Ground and Grounding}\label{sec:ja5} 

Common Ground and Grounding establish and continually confirm the knowledge and understanding that partners share, so that later action can rest on a common informational basis. In HRC, this process is comparatively procedural. In the coded literature, grounding is realized mainly through three recurring processes: \emph{Sharing}, \emph{Confirmation}, and \emph{Clarification}. These correspond to how information becomes mutually available, how sufficient understanding is confirmed, and how insufficient or mismatched understanding is resolved. Dynamic confirmation remains an open problem.

\paragraph{Sharing.}\label{sec:ja5-sharing}
Grounding a joint activity requires sharing task, environmental, and intentional information, primarily through shared representations and reciprocal exchange. Shared representations establish a common information basis accessible to both partners. Shared interfaces can externalize team states and action plans \cite{index14}, while shared artifacts allow partners to continuously negotiate common creative goals around the same object \cite{index45,index73}. This externalization goes beyond a specific referent (Section~\ref{sec:ja4}) to richer cognitive states. Shared representations and reciprocal exchange therefore serve complementary roles: the former provides a common basis for interaction, while the latter revises that basis as task-relevant information changes. Specifically, partners exchange information to interpret indirect requests \cite{index26}, communicate environmental changes that the user may not have noticed to maintain state alignment \cite{index51}, and use language, photographs, and LiDAR maps to establish remote spatial relations \cite{index96}.

\paragraph{Confirmation.}\label{sec:ja5-confirmation}
In HRC, confirmation is a checkpoint for whether to proceed or to pause. It occurs mainly at three levels. First, understanding confirmation: robots can indicate through verbal feedback that a user request has been understood, or seek further clarification when interpretation remains uncertain \cite{index5,index26,index187}. Second, action confirmation makes confirmation a condition for the next stage, such as resuming navigation after the user confirms \cite{index9}, accepting scan results \cite{index27}, completing an object handover \cite{index104}, or executing a risky behavior \cite{index143}. Third, agreement confirmation uses acceptance, rejection, or understanding feedback to verify whether partners have reached a shared solution or knowledge state \cite{index111,index129,index140}.
Explicit confirmation can ground reliably, but involving users in decisions increases coordination cost (e.g., delay and frustration) \cite{index105,index143}.
Across these implementations, confirmation connects a communicative response to a decision about task progression. Its placement determines what that response enables, while the requirement to obtain it introduces the coordination costs reported above.

\paragraph{Clarification.}\label{sec:ja5-clarification}
When communication is ambiguous, partners need to add or adjust information to restore a shared understanding they can act on. Clarification includes both immediate repair of the current ambiguity and the gradual formation of coordination conventions through repeated repair. Immediate repair can mean asking the user for additional pointing when a target is underspecified \cite{index21}, turning domain-specific language into verbal and embodied instructions that can be carried out \cite{index15}, or rewriting vague perspective expressions so that they are more explicit \cite{index109}. As interaction continues, these local repairs may settle into shared rules of expression. For example, partners may develop a common spatial language by repeatedly adjusting card positions \cite{index61}, or gradually select effective handover gestures from the robot's feedback \cite{index86}. Clarification therefore both restores the current collaboration and, through repeated grounding, reshapes future interaction by establishing partner-specific conventions.


\subsubsection{Task Co-representation}\label{sec:ja3} 
Task Co-representation concerns how each partner's contribution is represented as part of the larger joint task. Across the coded literature, existing work operationalizes this primarily as an explicit encoding of contribution dependencies: systems decompose the joint task into meaningful units, associate those units with roles and partner capabilities, and represent how one partner's action enables, constrains, or changes the other's possible contributions. Existing work can therefore be organized along four closely related representational targets: \emph{Task}, \emph{Role}, \emph{Partner}, and \emph{Action}. In most coded studies, these dependency representations are built into the robot's computational model; few studies check whether the human partner holds a compatible view of the same structure.

\paragraph{Task.}\label{sec:ja3-task}
Task representations organize interdependent contributions by specifying task units, dependencies, and contextual constraints. Task structure decomposes the overall task into recognizable units, such as functional modules \cite{index1}, sequential phases \cite{index104,index187}, and task steps \cite{index27}. Dependency rules connect these units by specifying when one step can follow another, for example by triggering subsequent actions from state changes \cite{index19} or determining execution order from dependencies among task elements \cite{index21}. Contextual constraints further specify which contributions are feasible under particular goals, partner capabilities, resource requirements, and environmental conditions, including task goals and agent skills \cite{index50}, future resource demands \cite{index105}, interaction history \cite{index146}, as well as safety constraints and domain-specific workflows \cite{index9,index15}.

\paragraph{Role.}\label{sec:ja3-role}
The joint task representation also encodes each partner’s authority, responsibilities, and collaborative position, mainly through task roles and relational roles. Task roles specify responsibilities, action authority, and decision boundaries within a specific task, for example by assigning responsibility to clearly defined tasks \cite{index14}, representing leadership and coordination through leader, follower, and collaborative roles \cite{index50}, or retaining decision authority with the human while delegating execution to the robot \cite{index61}. Relational roles instead capture how a partner is understood within the broader collaboration. The same robot, for example, may be perceived as an accompanist, teacher, performer, composer, or score, with these roles shifting over the course of interaction \cite{index47}.

\paragraph{Partner.}\label{sec:ja3-partner}
Existing studies support partner modeling through three approaches: behavioral inference, attribute modeling, and mental-state modeling. Behavioral inference derives expectations from observable partner behavior, such as bidding behavior \cite{index23}, interaction history \cite{index31,index148}, and task choices or temporal behavior \cite{index51,index114}. Attribute modeling builds expectations from relatively stable partner characteristics, including capability type \cite{index53}, bodily and motion characteristics \cite{index67}, and relative human--robot capabilities \cite{index71}. Mental-state modeling explicitly represents a partner’s internal understanding, including belief states \cite{index51}, intentions and strategies \cite{index148}, and perspectives or mistaken beliefs \cite{index109,index111}.

\paragraph{Action.}\label{sec:ja3-action}
Action representations situate individual actions within the dependency structure of the joint task—what an action requires, what it enables, and how it changes the other partner's possible contributions. Existing studies encode this through structured, probabilistic, and relational representations. Structured representations encode actions as explicit units, sequences, or states, including task-oriented grasping actions \cite{index12}, predefined motion sequences \cite{index18}, divided or alternating action structures \cite{index45}, and transitions between success and failure states \cite{index88}. Probabilistic representations capture possible subsequent actions through decision simulation \cite{index33}, path probabilities \cite{index40}, and task-rule constraints \cite{index134}. Relational representations further situate actions within a joint action structure, including shared action space \cite{index70}, contributions to a joint goal \cite{index117}, and the influence of one partner's action on the other's subsequent action space \cite{index140}.

\subsubsection{Shared Intention and Cooperative Commitment}\label{sec:ja2} 
Joint activity depends on sustained engagement from both partners. Shared Intention and Cooperative Commitment provide a shared direction for the joint task and support continued participation when demands change or difficulties arise. Across the coded studies, this commitment appeared in three recurring forms: \emph{Strategy Negotiation}, \emph{Mutual Responsiveness}, and \emph{Sustained Commitment}, corresponding to how a shared course of action is established, enacted through reciprocal response, and maintained over time. Most evidence in the coded literature is behavioral: studies show that partners adapt to each other or persist through difficulty, but rarely test whether these behaviors reflect shared intentional states or arise from task constraints, role obligations, or lack of alternatives.

\paragraph{Strategy Negotiation.}\label{sec:ja2-strategy}
When multiple task strategies are possible, humans and robots need to jointly establish or adjust a course of action. In collaborative transport, robots infer the current strategy from joint motion and reduce uncertainty between alternative routes \cite{index40}; in creative co-creation, partners progressively develop a shared direction through alternating contributions \cite{index24}. Uncertainty-aware control improves task success and perceived fluency \cite{index40}, while rapid alternation increases novelty density in co-creation \cite{index24}. These mechanisms transform local actions into evidence about a common course of action.

\paragraph{Mutual Responsiveness.}\label{sec:ja2-responsiveness}
One partner's ongoing actions, effort, or contributions can serve as input to the other's next action, whose response in turn shapes the subsequent joint activity. In teaching tasks, participants slow down and segment their demonstrations when the robot displays greater effort \cite{index129}; in co-creation, human and robot contributions form a rapid action--response loop \cite{index24}. Such contingency keeps the joint activity mutually shaped: each partner’s contribution responds to and reshapes the other’s. Behavioral adaptation can be stronger than participants' explicit self-reported awareness \cite{index129}. Conversely, when participants cannot determine whether the robot is responding to them, they perceive it as uncooperative and shift toward passive following \cite{index179}, highlighting the importance of both actual and perceived responsiveness.

\paragraph{Sustained Commitment.}\label{sec:ja2-commitment}
Even when a joint activity encounters failure, constraints, or changing task demands, partners need to remain invested in the activity. Robots can sustain collaboration through affective reassurance or cognitive guidance after failure \cite{index2}, or by displaying visible effort through slower and more segmented movements \cite{index129}. Human partners may sustain collaboration through troubleshooting and redistribution of responsibilities \cite{index14,index15,index102}, or by adapting their requests to the robot's capabilities \cite{index67,index109}. Affective and cognitive feedback improves perceived teamwork \cite{index2}, while greater robot effort elicits greater patience and effort from human partners \cite{index129}.

The common thread is that these representations organize what partners can do next: which target to act on, whether to proceed, how a contribution fits the joint task, and how to continue together. Prediction builds on this basis by asking what each partner is likely to do.

\subsection{Prediction}\label{sec:prediction}
Prediction concerns how partners use existing representations and ongoing action information to anticipate one another's behavior. Such expectations may arise from perceiving and interpreting a partner's unfolding actions (Section~\ref{sec:ja6}), as well as from knowledge of task roles, responsibilities, and complementary contributions (Section~\ref{sec:ja7}). At this level, the central issue is how understanding of a partner becomes expectations that allow each agent to prepare for and adapt to the other's behavior.

\subsubsection{Task Sharing and Complementary Contribution}\label{sec:ja7} 
Task Sharing and Complementary Contribution provides task-based expectations by constraining what each partner is likely to contribute. Where Task Co-representation (Section~\ref{sec:ja3}) addresses how the joint task structure is represented, task sharing addresses how the actual distribution of work reduces uncertainty about partner behavior. In the coded literature, division, leadership, and assistance address different dimensions of this uncertainty: division constrains what contribution to expect, leadership constrains whose action or judgment currently guides the interaction, and assistance specifies when the ordinary division of work may be supplemented or temporarily changed.

\paragraph{Division constrains what to expect.}\label{sec:ja7-division}
Division constrains the space of possible contributions. Many systems make partner behavior predictable by assigning complementary responsibilities in advance: humans retain decision making while robots execute \cite{index26,index27,index51,index92}, or robots provide information while humans perform the physical operation \cite{index147}. Allocation can also be adjusted when conflicts, failures, or task states change, for example by switching action targets \cite{index33,index41} or reallocating tasks after failures \cite{index34,index88}. Beyond allocation itself, handovers \cite{index12,index104}, turn-taking \cite{index61,index122}, and co-manipulation \cite{index22,index64,index68} create conditional relations between contributions: completion or initiation by one partner changes what the other is expected to do next. Each partner therefore reasons within an action space already narrowed by the division of labor.

\paragraph{Leadership constrains whose behavior to follow.}\label{sec:ja7-leadership}
Even when responsibilities are divided, a further question remains: whose direction, timing, or judgment should organize the next step. Knowing who currently leads tells each partner whose behavior to treat as the reference for coordination. Some studies fix this relation in advance, with either the human \cite{index36,index96} or the robot \cite{index136} leading. Others allow initiative to transfer explicitly through mode switching \cite{index9}, human takeover \cite{index50}, or yielding after conflict \cite{index98}. In still other cases, leadership emerges from information flow \cite{index22}, reaction speed \cite{index64}, or physical coupling \cite{index68}. Regardless of how it is established, leadership matters for prediction because it specifies which partner's behavior the other should currently anticipate and prepare for.

\paragraph{Assistance makes the contribution structure conditional.}\label{sec:ja7-assistance}
The expected division of work can also change when a partner lacks information, capability, time, or resources. Robots provide informational guidance \cite{index39}, cognitive suggestions \cite{index31,index148}, physical support \cite{index21}, or temporary task substitution when the human is unavailable \cite{index71}. Humans compensate by providing feedback \cite{index103}, supplying information unavailable to the robot \cite{index187}, or temporarily taking over when robot capability is insufficient or failures occur \cite{index16,index133}. More adaptive systems determine whether and when to help from the partner's strategy \cite{index31}, estimated availability and future task demands \cite{index114}, or current state \cite{index148}. Assistance thus adds contingency to task sharing: what a partner should contribute is no longer fixed by role alone, but depends on the evolving condition of the other partner and the task.

\subsubsection{Partner-Action Processing}\label{sec:ja6} 
Task structure can constrain what a partner is expected to do, but it cannot determine what the partner is actually doing at a given moment. Partner-Action Processing therefore maintains and revises expectations from ongoing evidence. Across the corpus, observation, action prediction, and monitoring are all common, but they function as mutually supporting processes: observations provide evidence about the current interaction, predictions project that evidence forward, and monitoring tests whether the resulting expectations remain compatible with subsequent behavior and task progress.

\paragraph{Observation selects evidence relative to the coordination problem.}\label{sec:ja6-observation}
What needs to be observed depends on how the partners are coupled. In physically coupled tasks, systems rely on hand and body pose \cite{index12,index62,index74}, trajectories \cite{index33,index40}, force direction \cite{index19,index20,index98}, shared-object motion \cite{index22,index36}, and handover contact \cite{index30,index88}. In less tightly coupled interaction, gaze \cite{index41}, verbal goals \cite{index21,index23}, spatial maps \cite{index96}, and workspace states \cite{index187} provide the context needed to interpret behavior. The relevant input is information selected relative to the coordination problem at hand.

\paragraph{Action prediction operates at multiple behavioral scales.}\label{sec:ja6-prediction}
Existing systems anticipate immediate motion and force \cite{index20,index50,index62,index92,index106}, higher-level action strategies \cite{index23,index31,index40}, and even whether a partner is likely to provide assistance \cite{index114}. These targets differ in temporal horizon and in what downstream decision they support. Anticipating a hand trajectory is useful only if it arrives early enough to modify motion; anticipating a strategy or willingness to help may instead inform task allocation or initiative. The value of an action prediction thus depends on whether it reaches the granularity and time horizon that the subsequent coordination decision requires, though few studies directly compare these factors.

\paragraph{Monitoring keeps expectations revisable.}\label{sec:ja6-monitoring}
Partners track whether unfolding behavior, coordination, and task progress remain consistent with current expectations. Existing systems monitor user responses and participation \cite{index103,index109,index71,index148}, conflicts and handover phases \cite{index33,index40,index74,index104}, agreement on joint solutions \cite{index111}, and task or system conditions such as confidence, path status, progress, and equipment state \cite{index1,index9,index114,index14,index15,index16,index102,index133}. Monitoring evaluates whether an existing expectation can still be relied upon, and grounds adaptation or, when deviations become consequential, repair (Section~\ref{sec:ja9}).

Better coordination can also emerge from interaction design. In one line of work comparing inference with explicit communication, multimodal explicitness compensated for imperfect prediction \cite{index92,index58}. More broadly, predictable action structures and recurring conventions can shift part of the predictive burden from inference to interaction design \cite{index28,index66,index73}. The design question is therefore how to distribute uncertainty: infer it from behavior, reduce it through task structure, or make it explicit through communication.

\subsection{Integration}\label{sec:integration}
Integration concerns how partners combine their actions and predicted effects to produce coordinated joint action. In HRC, integration is achieved both by reacting accurately to a partner and by making the coordination problem itself easier: robots make upcoming behavior readable, constrain when and how actions can unfold, and establish recurring behavioral regularities. When coordination nonetheless breaks down, recovery targets the specific joint relation that has failed---shared understanding, task organization, or execution coupling---each requiring a distinct repair process. Even when coordination and repair succeed, how participants experience control and authorship over the joint outcome remains a separate question.

\subsubsection{Action Coordination and Coordination Smoothers}\label{sec:ja8} 
Action Coordination concerns how otherwise separate contributions are made to work together in time, space, and ongoing interaction. A notable pattern in the coded literature is that many systems reshape the interaction to reduce the prediction required: they make upcoming actions easier to read, constrain when and how actions can unfold, or establish recurring behavioral regularities. Online synchronization and adaptation then handle the uncertainty that remains.

\paragraph{Reducing uncertainty through predictability.}\label{sec:ja8-predictability}
Robots frequently make their upcoming behavior available before coordination requires a response. Explicit signals include verbal requests and status information \cite{index26,index88,index133,index143,index147}, gaze and gesture \cite{index74,index86,index104}, AR and other visual feedback \cite{index41,index119}, and auditory or vibrotactile cues \cite{index28,index74}. In other cases, task actions themselves become signals: movement direction, force, rhythm, or trajectory provides information about what the robot is about to do \cite{index36,index40,index122}. Repeated interaction can further turn such regularities into conventions, including recurring action patterns \cite{index66}, turn-completion behaviors \cite{index73}, and melodic structures anticipating an upcoming handover \cite{index28}. These mechanisms reduce the amount of inference required from the partner. Human-perceptible behavioral patterns can improve predictability even when they depart from conventionally optimized plans \cite{index66}. The relevant design objective is sufficient predictability for the partner to prepare a complementary action.

\paragraph{Constraining the coordination problem.}\label{sec:ja8-constraining}
HRC systems also make coordination tractable by reducing the number of possible transitions that partners must negotiate online. Turns, task phases, and handover stages provide recognizable action boundaries \cite{index2,index12,index24,index30,index45,index61,index73,index74,index104,index131}, while transition rules determine when the interaction can advance through contact \cite{index9}, verbal confirmation \cite{index88}, user feedback \cite{index103,index105}, or risk conditions \cite{index143,index147,index187}; external constraints such as fixed locations and equipment cycles impose additional ordering \cite{index16,index102,index133}. Such structures transform an open-ended coordination problem into a smaller set of locally meaningful decisions. This can improve efficiency \cite{index27} and prevent conflicts \cite{index33}, but rigid protocols may increase coordination costs \cite{index5,index53,index105,index143} and constrain flexibility \cite{index12,index27}.

\paragraph{Handling residual uncertainty through online coupling.}\label{sec:ja8-coupling}
Where uncertainty cannot be removed in advance, partners coordinate by continuously coupling their ongoing behavior. Robots adjust speed \cite{index28}, acceleration \cite{index30}, movement frequency \cite{index136}, trajectory \cite{index18,index22,index40,index64}, force \cite{index20,index98}, and handover timing \cite{index30,index62,index104} in response to the human. Humans also change rhythm \cite{index80}, posture \cite{index179}, and exchange strategies \cite{index86} in response to robot behavior. At a broader level, teams may reorganize work around robot capabilities and limitations \cite{index16,index67}. Coupling tightness and coordination quality can diverge: behavioral synchronization does not reliably predict task success \cite{index22}, and signaling intended to facilitate coordination can disrupt ongoing human action when delivered at the wrong moment \cite{index137}. Field studies make a further issue visible: fluent robot operation can depend on troubleshooting, environmental preparation, and continuous adjustment by workers and supporting specialists \cite{index14,index15,index16,index102}, so that what appears as smooth coordination may reflect substantial invisible work by the human side of the team. These field studies have identified such work; how technical evaluations of HRC systems can account for maintenance and preparation costs that fall outside the measured task remains an open question.

\subsubsection{Breakdown Detection and Repair}\label{sec:ja9} 
Breakdown Detection and Repair encompasses distinct recovery procedures, because joint activity can fail at different relations between the partners. The same observable failure may require fundamentally different recovery depending on whether the breakdown concerns what the partners understand, how the joint activity is organized, or whether an otherwise valid course of action can still be executed.

\paragraph{Repairing shared understanding.}\label{sec:ja9-epistemic}
Some breakdowns are epistemic: partners no longer share a usable interpretation of the relevant object, state, instruction, or intended action. Recovery involves clarification, reformulation, renewed reference, or additional evidence of understanding. Robots may clarify ambiguous instructions \cite{index5,index26}, make perspective-dependent expressions more explicit \cite{index109}, communicate changes that invalidate the human's previous understanding \cite{index51}, or re-establish spatial grounding after a mismatch \cite{index61,index96}. Corrections and negotiation can also be used to determine whether a revised interpretation is accepted \cite{index111,index147}. Communication by itself is not repair; it becomes repair when it restores an understanding from which the joint activity can proceed.

\paragraph{Repairing plans and contribution structures.}\label{sec:ja9-organizational}
Other breakdowns leave the relevant objects and states understood but invalidate how the joint activity has been organized. Conflicting actions can trigger negotiation of alternative choices \cite{index33,index111}; infeasible routes can be revised \cite{index50,index96}; failed subtasks can be reassigned \cite{index88}; and control or leader--follower relations can change when the original coordination strategy no longer works \cite{index20,index98,index133,index179}. Recovery in these cases reorganizes future contributions.

\paragraph{Repairing execution.}\label{sec:ja9-execution}
A third class of breakdown occurs when the plan remains usable but its execution has failed. Partners can repeat an action or handover \cite{index66,index104,index129,index146}, return to a previous stable state through recalibration or rollback \cite{index14,index15,index27,index41}, or re-establish the physical or temporal conditions required for the original course of action \cite{index88}. The object of repair here is the execution coupling through which already-compatible contributions are carried out.

This distinction reveals an important property of Integration: repair is recursive rather than terminal. A breakdown detected during action may require partners to revisit a shared reference, update common ground, or revise the task structure, while representational mismatches may become visible only when partners attempt to act. Breakdown and repair therefore provide concrete evidence that Representation, Prediction, and Integration are not a one-way sequence. An open problem is the diagnosis that precedes repair: many recovery mechanisms can respond once the relevant failure is known, some systems can localize internal faults and decide when to involve the human \cite{index1}, but determining which joint relation between the partners has failed, and confirming that the repaired state is again mutually usable, is rarely treated as a collaborative process in its own right.

\subsubsection{Sense of Individual and Joint Agency}\label{sec:ja10} 
Successful coordination leaves open how the resulting action is experienced. Sense of Individual and Joint Agency concerns whether participants experience or attribute control, authorship, and ownership of joint effects to themselves, the partner, or the acting pair. This distinction matters in HRC because objective control allocation---who can modify the robot's motion, who decides whether to proceed---is straightforward to manipulate, yet how control is experienced remains a separate question. A human may retain formal decision authority without experiencing authorship over the unfolding joint action, while a robot may make substantial contributions without being treated as a co-author of the outcome.

The clearest evidence in the corpus concerns cases in which attribution itself is examined. Relative robot status changes how participants assign responsibility for collaborative outcomes \cite{index188}; cooperative reaction tasks show that human behavior can become organized around a robot as a co-agent under particular interaction conditions \cite{index70,index80}. In co-creative settings, participants shift from ``my drawing'' to ``our creation'' as robot contributions become more salient \cite{index73}, while the same observable robot contribution can be interpreted as autonomous initiative, compliant execution, or part of a jointly produced action \cite{index73}. Experienced agency therefore diverges from the objective distribution of control.

Although many coded studies involve control allocation, autonomy levels, or leader--follower relations (Appendix, Table~\ref{tab:app_detailed_codebook}), substantially fewer directly measure whether participants experience self-agency, shared agency, or ownership of the joint outcome. Studies of autonomy level, leader--follower relations, takeover, or task performance provide relevant antecedents, but measuring agency requires assessing experienced control or authorship directly. The open question is how to pair objective coordination and control measures with direct measures of experienced agency, particularly in systems where robot initiative and adaptation make the source of a joint outcome increasingly difficult to attribute.

%% file: sections/5_discussion.tex
\section{Literature Analysis and Findings}












\subsection{Analysis Method and Result}
\begin{figure*}[tbp]
    \centering
    \includegraphics[width=\textwidth,height=0.78\textheight,keepaspectratio]{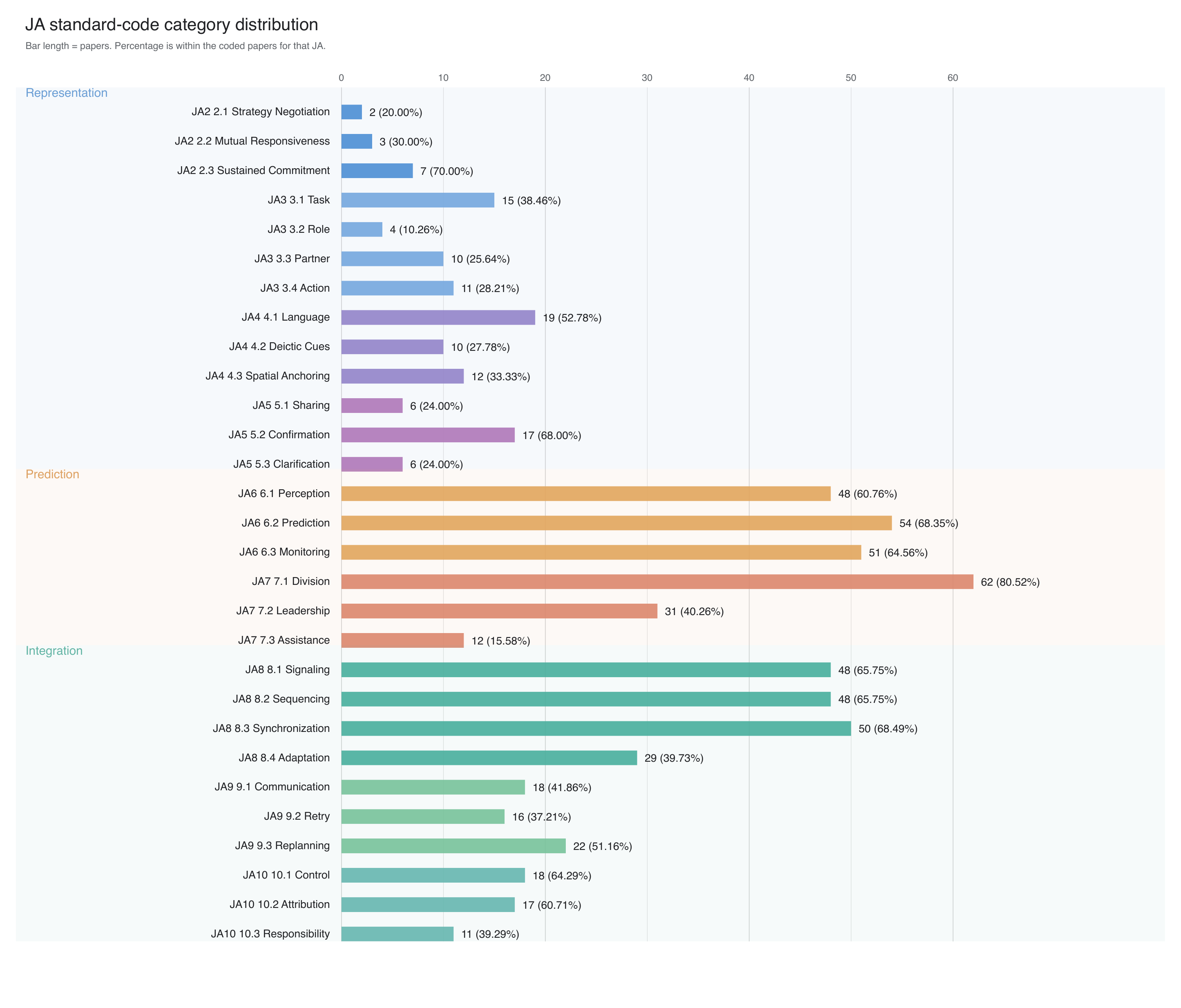}
    \caption{Distribution of coded categories within JA2--JA10, grouped into Representation, Prediction, and Integration. Bar lengths indicate the number of papers assigned to each category; percentages use the papers coded for the corresponding JA construct as the denominator. Categories are non-exclusive, so percentages within a construct can sum to more than 100\%.}
    \Description{Horizontal bar chart of 29 categories grouped by nine Joint Action constructs and three layers. Each bar reports a paper count and its percentage within the corresponding construct.}
    \label{fig:ja-category-distribution}
\end{figure*}

For the given 83 papers and the corresponding open coding result, we first conducted frequency analysis for capacities (Representation, Prediction, Integration), JA (JA2--JA10), and subcodes (Table~\ref{tab:category_distribution} and Figure~\ref{fig:ja-category-distribution}). At the capacity level, \textit{Representation} is less frequent than the other two capacities, covering 72.3\% of the reviewed papers. At the JA level, JA6, JA7, and JA8 are substantially more frequent than the other categories, whereas JA2 is comparatively much less frequent. At the subcode level, \textit{3.2 Role} and \textit{7.3 Assistance} occur less frequently than their respective parent JAs.

\begin{figure*}[tbp]
    \centering
    \includegraphics[width=\textwidth,height=0.78\textheight,keepaspectratio]{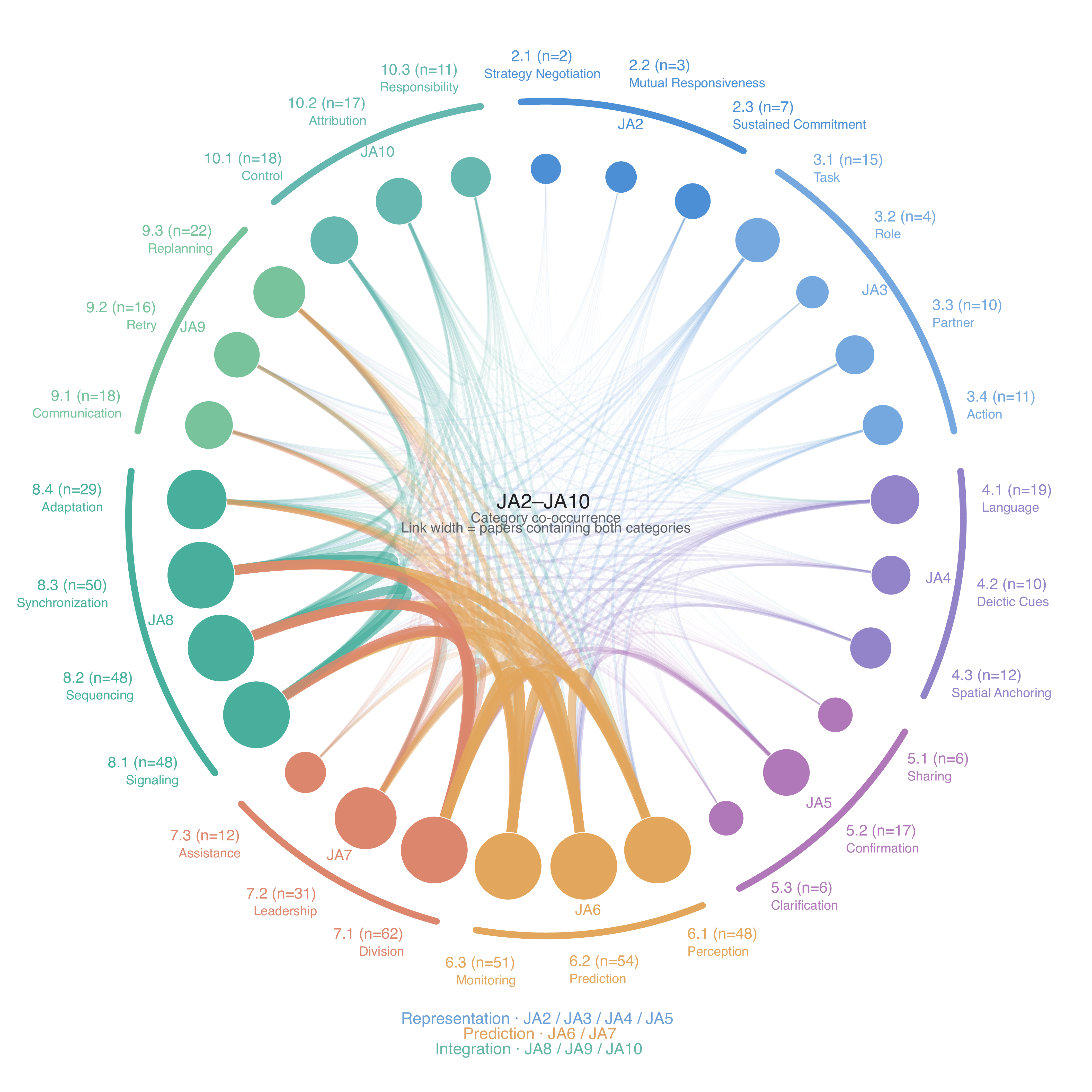}
    \caption{Paper-level co-occurrence of coded categories within JA2--JA10. Categories are arranged by construct and grouped into Representation, Prediction, and Integration. Labels report the number of papers assigned to each category, and link widths indicate the number of papers containing both linked categories.}
    \Description{Circular diagram connecting 29 coded categories across Representation, Prediction, and Integration. Links show category co-occurrence within papers, with wider links representing more shared papers.}
    \label{fig:ja-category-cooccurrence}
\end{figure*}

We further analyze the co-occurrence patterns among subcodes (Figure~\ref{fig:ja-category-cooccurrence}), where the thicker and more opaque lines indicate that a pair of subcodes co-occurs in a larger number of papers, whereas thinner and more transparent lines indicate less frequent co-occurrence.
To better examine the relationship between subcodes, we introduce the \textit{lift} metric (Equation~\ref{equ:lift}). A lift value of 1 indicates that two (or three) subcodes occur independently. A value greater than 1 indicates that they co-occur more frequently than would be expected under independence, while a value below 1 indicates that they co-occur less frequently than expected.
We also introduce the \textit{independent expectation} metric (Equation~\ref{equ:independent_exp}), which is the expected co-occurrence count under independence. Comparing it with the observed count shows whether a combination of subcodes co-occurs more or less frequently than expected.
We use \textit{lift} and \textit{independent expectation} to filter the co-occurrence data, especially those whose \textit{independent expectation} is at least 4 and $|\text{Observed} - \text{Expected}| \geq 4$~\footnote{This threshold is calculated as $\lfloor 83 \times 0.05 \rfloor$, where 83 is the total number of papers. We introduce this threshold because some subcode combinations have very low expected co-occurrence counts. In such cases, a difference of only one or two papers from the expected count can still produce a relatively large lift value (e.g., 2--3). Relying only on extreme lift values would therefore disproportionately highlight these rare combinations, even though such small absolute differences may have limited practical significance.}. Table~\ref{tab:cooccurrence_patterns} illustrates the filtered result.

To further examine the dependency among subcodes, we conducted an additional analysis. Specifically, we calculated the proportion of papers containing a given subcode combination relative to the number of papers containing each individual subcode. This allows us to identify cases where the presence of one subcode is strongly associated with the presence of another. Based on these proportions, we identified several pairs and triplets of subcodes that exhibit particularly strong mutual dependencies (Table~\ref{tab:dependency_patterns}). For pairs, the table only includes combinations for which at least one conditional proportion exceeds 60\%; for triplets, it only includes combinations for which at least one conditional proportion exceeds 30\%.
We also identify three dominating subcodes that exhibit strong unidirectional dependencies with multiple other subcodes in Table~\ref{tab:directional_dependency}.

\begin{table*}[t]
    \centering
    \caption{Distribution of categories.}
    \label{tab:category_distribution}
    \small
    \begin{tabular}{llrlr}
    \toprule
    \textbf{Level} & \textbf{Category} & \textbf{Papers} & \textbf{Broad Group} & \textbf{Percent} \\
    \midrule

    Capacity & Prediction     & 83 & --             & 100.0\% \\
    Capacity & Integration    & 81 & --             & 97.6\% \\
    Capacity & Representation & 60 & --             & 72.3\% \\

    \midrule
    JA & JA6  & 79 & Prediction     & 95.2\% \\
    JA & JA7  & 77 & Prediction     & 92.8\% \\
    JA & JA8  & 73 & Integration    & 88.0\% \\
    JA & JA9  & 43 & Integration    & 51.8\% \\
    JA & JA3  & 39 & Representation & 47.0\% \\
    JA & JA4  & 36 & Representation & 43.4\% \\
    JA & JA10 & 28 & Integration    & 33.7\% \\
    JA & JA5  & 25 & Representation & 30.1\% \\
    JA & JA2  & 10 & Representation & 12.0\% \\

    \midrule
    Subcode & 7.1 Division                 & 62 & Prediction     & 74.7\% \\
    Subcode & 6.2 Prediction               & 54 & Prediction     & 65.1\% \\
    Subcode & 6.3 Monitoring               & 51 & Prediction     & 61.4\% \\
    Subcode & 8.3 Synchronization          & 50 & Integration    & 60.2\% \\
    Subcode & 6.1 Perception               & 48 & Prediction     & 57.8\% \\
    Subcode & 8.1 Signaling                & 48 & Integration    & 57.8\% \\
    Subcode & 8.2 Sequencing               & 48 & Integration    & 57.8\% \\
    Subcode & 7.2 Leadership               & 31 & Prediction     & 37.3\% \\
    Subcode & 8.4 Adaptation               & 29 & Integration    & 34.9\% \\
    Subcode & 9.3 Replanning               & 22 & Integration    & 26.5\% \\
    Subcode & 4.1 Language                 & 19 & Representation & 22.9\% \\
    Subcode & 10.1 Control                 & 18 & Integration    & 21.7\% \\
    Subcode & 9.1 Communication            & 18 & Integration    & 21.7\% \\
    Subcode & 10.2 Attribution             & 17 & Integration    & 20.5\% \\
    Subcode & 5.2 Confirmation             & 17 & Representation & 20.5\% \\
    Subcode & 9.2 Retry                    & 16 & Integration    & 19.3\% \\
    Subcode & 3.1 Task                     & 15 & Representation & 18.1\% \\
    Subcode & 4.3 Spatial Anchoring        & 12 & Representation & 14.5\% \\
    Subcode & 7.3 Assistance               & 12 & Prediction     & 14.5\% \\
    Subcode & 10.3 Responsibility          & 11 & Integration    & 13.3\% \\
    Subcode & 3.4 Action                   & 11 & Representation & 13.3\% \\
    Subcode & 3.3 Partner                  & 10 & Representation & 12.0\% \\
    Subcode & 4.2 Deictic Cues             & 10 & Representation & 12.0\% \\
    Subcode & 2.3 Sustained Commitment     & 7  & Representation & 8.4\% \\
    Subcode & 5.1 Sharing                  & 6  & Representation & 7.2\% \\
    Subcode & 5.3 Clarification            & 6  & Representation & 7.2\% \\
    Subcode & 3.2 Role                     & 4  & Representation & 4.8\% \\
    Subcode & 2.2 Mutual Responsiveness    & 3  & Representation & 3.6\% \\
    Subcode & 2.1 Strategy Negotiation     & 2  & Representation & 2.4\% \\

    \bottomrule
    \end{tabular}
\end{table*}

\begin{equation}
\begin{gathered}
    \operatorname{Lift}(A,B)
    =
    \frac{P(A \cap B)}{P(A)\,P(B)},\\
    \operatorname{Lift}(A,B,C)
    =
    \frac{P(A \cap B \cap C)}
    {P(A)\,P(B)\,P(C)}
\end{gathered}
    \label{equ:lift}
\end{equation}

\begin{equation}
\begin{gathered}
    E[n(A \cap B)]
    =
    N \cdot P(A)\,P(B)
    =
    \frac{n(A)\,n(B)}{N},\\
    E[n(A \cap B \cap C)]
    =
    N \cdot P(A)\,P(B)\,P(C)
    =
    \frac{n(A)\,n(B)\,n(C)}{N^2}
\end{gathered}
    \label{equ:independent_exp}
\end{equation}

\begin{table*}[t]
    \centering
    \caption{Filtered co-occurrence patterns among subcodes.}
    \label{tab:cooccurrence_patterns}
    \small
    \resizebox{\textwidth}{!}{
    \begin{tabular}{llrrrcccc}
    \toprule
    \textbf{Scope} &
    \textbf{Capacity Combination} &
    \textbf{Subcode 1} &
    \textbf{Subcode 2} &
    \textbf{Subcode 3} &
    \textbf{Observed} &
    \textbf{Expected} &
    \textbf{Obs. $-$ Exp.} &
    \textbf{Lift} \\
    \midrule

    Pair within &
    Integration $\times$ Integration &
    8.1 Signaling &
    8.3 Synchronization &
    -- & 35 & 28.92 & 6.08 & 1.21 \\

    Pair within &
    Prediction $\times$ Prediction &
    6.1 Perception &
    6.3 Monitoring &
    -- & 24 & 29.49 & -5.49 & 0.81 \\

    Pair within &
    Prediction $\times$ Prediction &
    6.1 Perception &
    7.2 Leadership &
    -- & 24 & 17.93 & 6.07 & 1.34 \\

    Pair within &
    Integration $\times$ Integration &
    10.1 Control &
    10.2 Attribution &
    -- & 9 & 3.69 & 5.31 & 2.44 \\

    Pair within &
    Integration $\times$ Integration &
    8.3 Synchronization &
    9.1 Communication &
    -- & 5 & 10.84 & -5.84 & 0.46 \\

    \midrule

    Pair cross &
    Prediction $\times$ Integration &
    6.1 Perception &
    8.3 Synchronization &
    -- & 36 & 28.92 & 7.08 & 1.25 \\

    Pair cross &
    Prediction $\times$ Integration &
    6.2 Prediction &
    8.1 Signaling &
    -- & 27 & 31.23 & -4.23 & 0.86 \\

    Pair cross &
    Prediction $\times$ Integration &
    6.3 Monitoring &
    8.3 Synchronization &
    -- & 26 & 30.72 & -4.72 & 0.85 \\

    Pair cross &
    Prediction $\times$ Integration &
    6.3 Monitoring &
    9.3 Replanning &
    -- & 19 & 13.52 & 5.48 & 1.41 \\

    Pair cross &
    Prediction $\times$ Integration &
    6.3 Monitoring &
    9.2 Retry &
    -- & 14 & 9.83 & 4.17 & 1.42 \\

    Pair cross &
    Prediction $\times$ Integration &
    7.2 Leadership &
    9.3 Replanning &
    -- & 14 & 8.22 & 5.78 & 1.70 \\

    Pair cross &
    Integration $\times$ Prediction &
    10.1 Control &
    7.2 Leadership &
    -- & 13 & 6.72 & 6.28 & 1.93 \\

    Pair cross &
    Representation $\times$ Integration &
    4.1 Language &
    8.3 Synchronization &
    -- & 7 & 11.45 & -4.45 & 0.61 \\

    Pair cross &
    Prediction $\times$ Integration &
    6.1 Perception &
    9.2 Retry &
    -- & 4 & 9.25 & -5.25 & 0.43 \\

    Pair cross &
    Representation $\times$ Integration &
    3.3 Partner &
    8.1 Signaling &
    -- & 1 & 5.78 & -4.78 & 0.17 \\

    Pair cross &
    Representation $\times$ Integration &
    3.3 Partner &
    8.3 Synchronization &
    -- & 1 & 6.02 & -5.02 & 0.17 \\

    Pair cross &
    Representation $\times$ Prediction &
    3.3 Partner &
    6.1 Perception &
    -- & 1 & 5.78 & -4.78 & 0.17 \\

    Pair cross &
    Representation $\times$ Prediction &
    2.3 Sustained Commitment &
    6.1 Perception &
    -- & 0 & 4.05 & -4.05 & 0.00 \\

    \midrule

    Triplets within &
    Prediction $\times$ Prediction $\times$ Prediction &
    6.1 Perception &
    6.2 Prediction &
    7.1 Division &
    19 & 23.33 & -4.33 & 0.81 \\

    Triplets within &
    Prediction $\times$ Prediction $\times$ Prediction &
    6.1 Perception &
    6.3 Monitoring &
    7.1 Division &
    18 & 22.03 & -4.03 & 0.82 \\

    Triplets within &
    Prediction $\times$ Prediction $\times$ Prediction &
    6.1 Perception &
    6.2 Prediction &
    7.2 Leadership &
    16 & 11.66 & 4.34 & 1.37 \\

    Triplets within &
    Integration $\times$ Integration $\times$ Integration &
    10.1 Control &
    10.2 Attribution &
    8.3 Synchronization &
    7 & 2.22 & 4.78 & 3.15 \\

    \midrule

    Triplets RPI &
    Representation $\times$ Prediction $\times$ Integration &
    4.2 Deictic Cues &
    7.1 Division &
    8.2 Sequencing &
    9 & 4.32 & 4.68 & 2.08 \\

    Triplets RPI &
    Representation $\times$ Prediction $\times$ Integration &
    5.2 Confirmation &
    6.3 Monitoring &
    9.2 Retry &
    7 & 2.01 & 4.99 & 3.48 \\

    Triplets RPI &
    Representation $\times$ Prediction $\times$ Integration &
    5.2 Confirmation &
    7.1 Division &
    9.1 Communication &
    7 & 2.75 & 4.25 & 2.54 \\

    Triplets RPI &
    Representation $\times$ Prediction $\times$ Integration &
    5.2 Confirmation &
    7.1 Division &
    9.2 Retry &
    7 & 2.45 & 4.55 & 2.86 \\

    Triplets RPI &
    Integration $\times$ Representation $\times$ Prediction &
    10.3 Responsibility &
    4.1 Language &
    6.2 Prediction &
    6 & 1.64 & 4.36 & 3.66 \\

    \bottomrule
    \end{tabular}
    }
\end{table*}

\begin{table*}[t]
    \centering
    \caption{Subcode combinations with strong conditional dependencies.}
    \label{tab:dependency_patterns}
    \small
    \resizebox{\textwidth}{!}{
    \begin{tabular}{lllllcccc}
    \toprule
    \textbf{Scope} &
    \textbf{Capacity Combination} &
    \textbf{Subcode 1} &
    \textbf{Subcode 2} &
    \textbf{Subcode 3} &
    $\mathbf{P(\text{Combination}\mid\text{Subcode 1})}$ &
    $\mathbf{P(\text{Combination}\mid\text{Subcode 2})}$ &
    $\mathbf{P(\text{Combination}\mid\text{Subcode 3})}$ &
    \textbf{Lift} \\
    \midrule

    Pair cross &
    Prediction $\times$ Integration &
    6.1 Perception &
    8.3 Synchronization &
    -- &
    75.0\% &
    72.0\% &
    -- &
    1.245 \\

    Pair within &
    Integration &
    8.1 Signaling &
    8.3 Synchronization &
    -- &
    72.9\% &
    70.0\% &
    -- &
    1.210 \\

    Pair within &
    Prediction &
    6.2 Prediction &
    6.3 Monitoring &
    -- &
    66.7\% &
    70.6\% &
    -- &
    1.085 \\

    Pair within &
    Integration &
    8.1 Signaling &
    8.2 Sequencing &
    -- &
    64.6\% &
    64.6\% &
    -- &
    1.117 \\

    Pair within &
    Prediction &
    6.3 Monitoring &
    7.1 Division &
    -- &
    78.4\% &
    64.5\% &
    -- &
    1.050 \\

    Pair cross &
    Prediction $\times$ Integration &
    7.1 Division &
    8.2 Sequencing &
    -- &
    62.9\% &
    81.3\% &
    -- &
    1.088 \\

    Pair within &
    Prediction &
    6.2 Prediction &
    7.1 Division &
    -- &
    70.4\% &
    61.3\% &
    -- &
    0.942 \\

    Pair cross &
    Prediction $\times$ Integration &
    7.1 Division &
    8.1 Signaling &
    -- &
    61.3\% &
    79.2\% &
    -- &
    1.060 \\

    Pair cross &
    Prediction $\times$ Integration &
    6.3 Monitoring &
    8.2 Sequencing &
    -- &
    60.8\% &
    64.6\% &
    -- &
    1.051 \\

    Pair cross &
    Prediction $\times$ Integration &
    6.1 Perception &
    8.1 Signaling &
    -- &
    60.4\% &
    60.4\% &
    -- &
    1.045 \\

    \midrule

    Triplets within &
    Prediction &
    6.2 Prediction &
    6.3 Monitoring &
    7.1 Division &
    48.1\% &
    51.0\% &
    41.9\% &
    1.049 \\

    Triplets within &
    Integration &
    8.1 Signaling &
    8.2 Sequencing &
    8.3 Synchronization &
    41.7\% &
    41.7\% &
    40.0\% &
    1.196 \\

    Triplets within &
    Prediction &
    6.1 Perception &
    6.2 Prediction &
    7.1 Division &
    39.6\% &
    35.2\% &
    30.6\% &
    0.814 \\

    \bottomrule
    \end{tabular}
    }
\end{table*}

\begin{table*}[t]
    \centering
    \caption{Directional dependency patterns among subcodes.}
    \label{tab:directional_dependency}
    \small
    \resizebox{\textwidth}{!}{
    \begin{tabular}{llllccc}
    \toprule
    \textbf{Scope} &
    \textbf{Capacity Combination} &
    \textbf{Subcode 1} &
    \textbf{Subcode 2} &
    $\mathbf{P(\text{Combination}\mid\text{Subcode 1})}$ &
    $\mathbf{P(\text{Combination}\mid\text{Subcode 2})}$ &
    \textbf{Dependency} \\
    \midrule

    Pair cross &
    Representation $\times$ Prediction &
    7.1 Division &
    4.2 Deictic Cues &
    16.1\% &
    100.0\% &
    4.2 Deictic Cues $\rightarrow$ 7.1 Division \\

    Pair cross &
    Representation $\times$ Prediction &
    7.1 Division &
    2.3 Sustained Commitment &
    11.3\% &
    100.0\% &
    2.3 Sustained Commitment $\rightarrow$ 7.1 Division \\

    Pair cross &
    Representation $\times$ Prediction &
    7.1 Division &
    5.3 Clarification &
    9.7\% &
    100.0\% &
    5.3 Clarification $\rightarrow$ 7.1 Division \\

    Pair cross &
    Prediction $\times$ Integration &
    7.1 Division &
    9.1 Communication &
    27.4\% &
    94.4\% &
    9.1 Communication $\rightarrow$ 7.1 Division \\

    Pair cross &
    Representation $\times$ Prediction &
    7.1 Division &
    5.2 Confirmation &
    25.8\% &
    94.1\% &
    5.2 Confirmation $\rightarrow$ 7.1 Division \\

    Pair cross &
    Representation $\times$ Prediction &
    7.1 Division &
    4.3 Spatial Anchoring &
    17.7\% &
    91.7\% &
    4.3 Spatial Anchoring $\rightarrow$ 7.1 Division \\

    Pair cross &
    Representation $\times$ Prediction &
    7.1 Division &
    5.1 Sharing &
    8.1\% &
    83.3\% &
    5.1 Sharing $\rightarrow$ 7.1 Division \\

    Pair cross &
    Representation $\times$ Prediction &
    7.1 Division &
    3.4 Action &
    14.5\% &
    81.8\% &
    3.4 Action $\rightarrow$ 7.1 Division \\

    Pair cross &
    Prediction $\times$ Integration &
    7.1 Division &
    9.2 Retry &
    21.0\% &
    81.3\% &
    9.2 Retry $\rightarrow$ 7.1 Division \\

    Pair cross &
    Representation $\times$ Prediction &
    7.1 Division &
    3.1 Task &
    19.4\% &
    80.0\% &
    3.1 Task $\rightarrow$ 7.1 Division \\

    Pair cross &
    Prediction $\times$ Integration &
    7.1 Division &
    10.1 Control &
    22.6\% &
    77.8\% &
    10.1 Control $\rightarrow$ 7.1 Division \\

    Pair cross &
    Prediction $\times$ Integration &
    7.1 Division &
    10.2 Attribution &
    21.0\% &
    76.5\% &
    10.2 Attribution $\rightarrow$ 7.1 Division \\

    \midrule

    Pair cross &
    Representation $\times$ Prediction &
    6.3 Monitoring &
    3.3 Partner &
    17.6\% &
    90.0\% &
    3.3 Partner $\rightarrow$ 6.3 Monitoring \\

    Pair cross &
    Prediction $\times$ Integration &
    6.3 Monitoring &
    9.2 Retry &
    27.5\% &
    87.5\% &
    9.2 Retry $\rightarrow$ 6.3 Monitoring \\

    Pair cross &
    Prediction $\times$ Integration &
    6.3 Monitoring &
    9.3 Replanning &
    37.3\% &
    86.4\% &
    9.3 Replanning $\rightarrow$ 6.3 Monitoring \\

    Pair cross &
    Representation $\times$ Prediction &
    6.3 Monitoring &
    2.3 Sustained Commitment &
    11.8\% &
    85.7\% &
    2.3 Sustained Commitment $\rightarrow$ 6.3 Monitoring \\

    Pair within &
    Prediction &
    6.3 Monitoring &
    7.3 Assistance &
    19.6\% &
    83.3\% &
    7.3 Assistance $\rightarrow$ 6.3 Monitoring \\

    Pair cross &
    Representation $\times$ Prediction &
    6.3 Monitoring &
    5.3 Clarification &
    9.8\% &
    83.3\% &
    5.3 Clarification $\rightarrow$ 6.3 Monitoring \\

    Pair cross &
    Representation $\times$ Prediction &
    6.3 Monitoring &
    5.2 Confirmation &
    27.5\% &
    82.4\% &
    5.2 Confirmation $\rightarrow$ 6.3 Monitoring \\

    Pair cross &
    Prediction $\times$ Integration &
    6.3 Monitoring &
    9.1 Communication &
    27.5\% &
    77.8\% &
    9.1 Communication $\rightarrow$ 6.3 Monitoring \\

    Pair cross &
    Representation $\times$ Prediction &
    6.3 Monitoring &
    4.3 Spatial Anchoring &
    17.6\% &
    75.0\% &
    4.3 Spatial Anchoring $\rightarrow$ 6.3 Monitoring \\

    \midrule

    Pair cross &
    Representation $\times$ Prediction &
    6.2 Prediction &
    3.3 Partner &
    16.7\% &
    90.0\% &
    3.3 Partner $\rightarrow$ 6.2 Prediction \\

    Pair cross &
    Prediction $\times$ Integration &
    6.2 Prediction &
    10.1 Control &
    27.8\% &
    83.3\% &
    10.1 Control $\rightarrow$ 6.2 Prediction \\

    Pair cross &
    Representation $\times$ Prediction &
    6.2 Prediction &
    5.1 Sharing &
    9.3\% &
    83.3\% &
    5.1 Sharing $\rightarrow$ 6.2 Prediction \\

    Pair cross &
    Prediction $\times$ Integration &
    6.2 Prediction &
    10.3 Responsibility &
    16.7\% &
    81.8\% &
    10.3 Responsibility $\rightarrow$ 6.2 Prediction \\

    Pair cross &
    Representation $\times$ Prediction &
    6.2 Prediction &
    4.1 Language &
    27.8\% &
    78.9\% &
    4.1 Language $\rightarrow$ 6.2 Prediction \\

    Pair cross &
    Representation $\times$ Prediction &
    6.2 Prediction &
    5.2 Confirmation &
    24.1\% &
    76.5\% &
    5.2 Confirmation $\rightarrow$ 6.2 Prediction \\

    \bottomrule
    \end{tabular}
    }
\end{table*}

%% file: sections/6_theory.tex




%% file: sections/7_future.tex

\section{Discussion}
\subsection{Preference vs.\ Accommodation in Partner Modeling}
When robots infer human preferences from observed behavior, they cannot distinguish preference from accommodation to the robot's limitations.
Studies in Sec.~\ref{sec:ja2-responsiveness} and Sec.~\ref{sec:ja8-coupling} document how humans adapt their pace, posture, and work arrangement to robot limitations, while Sec.~\ref{sec:ja3-partner} shows that the robot's model of its human partner infers behavioral tendencies from interaction history. The same behavior can therefore be read as a personal preference or as an adaptation to the robot's constraints \cite{index179,index16,index67}. Existing mutual-adaptation models infer human adaptation and adjust robot strategy, but the evidence does not yet address whether, once a limitation is lifted, the system still retains the prior accommodation as a long-term preference \cite{Nikolaidis2017MutualAdaptation}.
This confound shapes long-term task allocation: workaround behavior recorded as preference becomes the basis for future assignments, keeping work on the human side that could be returned to the robot. When the system then updates its model from the division it imposed, behavioral predictions may grow more accurate while merely consolidating a work pattern shaped by its own limitations.
Future studies can introduce a robot limitation, then remove it and compare whether the system updates its model of the human or continues to treat the old workaround as a preference. Evaluation should track whether task assignments actually shift back to the robot and how much unnecessary human effort remains.

\subsection{Sense of Agency Beyond Action-Level Feedback}
Existing HRC studies assess sense of agency primarily at the action level, measuring whether the human can execute the current movement accurately. Whether the human understands the overall task state, anticipates what comes next, and judges when intervention is needed remains largely untested.
Sec.~\ref{sec:ja10} distinguishes control allocation from experienced agency, and Sec.~\ref{sec:ja3} notes that task dependencies are encoded in robot models while whether humans form a compatible understanding of those dependencies receives little direct testing. Evidence from shared-control studies illustrates the gap: changing auditory feedback alone raised agency ratings at the same automation ratio, indicating that current agency measures respond to sensory cues rather than task understanding \cite{Morita2022Agency}. Smooth action execution can therefore coexist with poor task-level grasp: a human may perform each step correctly yet misjudge the precondition for the next step or miss the window for redirecting the overall plan.
The core issue is that experiencing agency is not the same as being able to intervene effectively, and having control authority is not the same as being able to use it in time. Future work should measure both: experienced agency and cognitive task understanding, that is, whether the human knows the current task state, foresees what happens without intervention, and can actually alter the task trajectory within the available window \cite{Endsley1995SituationAwareness}. Pairing these measures will determine whether feedback that improves experienced agency also improves the ability to intervene at the right moment.

\subsection{Self-Improving Robots through Repair}
Current repair in HRC fixes the immediate breakdown but does not change the robot's future behavior, so the same human effort recurs every time the problem returns.
Sec.~\ref{sec:ja9} distinguishes repair at the understanding, task-organization, and execution levels, while field studies in Sec.~\ref{sec:ja8-coupling} show that recovery and smooth operation depend on human preparation, troubleshooting, and handover labor \cite{index14,index15}. These studies identify the value of preserving preparation history and handover knowledge; what remains to be established is how repair experience changes subsequent behavior and whether that change reduces the extra effort humans spend across events.
Research on organizational learning in hospitals offers a useful distinction: fixing the immediate disruption so work can continue versus identifying and addressing the underlying cause so the same disruption does not recur \cite{TuckerEdmondson2003Learning}. If a robot only learns to request the same kind of help faster, single-recovery time may shrink while the human's repeated preparation, explanation, and takeover persist; if it replicates a past repair under changed conditions, experience reuse can introduce new mismatches.
Future work can organize repair records into failure cause, applicable conditions, actions taken, and verified outcomes, and let the robot update task preconditions, execution strategies, or help-request rules accordingly. Comparing a system that stores only events and steps against one that stores context-dependent knowledge---which conditions the repair applies to and when it should not be reused---under both recurrence of the same problem and change in applicable conditions, will test the ability to prevent, reuse, and stop reusing repairs. Evaluation should accumulate the extra human effort per participant, recurrence of avoidable problems, and knowledge-maintenance costs, so that continuous system improvement maps onto actual reductions in team workload.

\subsection{Co-Training Transfer on Task Dependencies}
Human--robot co-training studies have not yet tested whether collaboration improvements transfer when task dependencies change, so it remains unclear what reusable knowledge each party actually acquires. Sec.~\ref{sec:ja3-task} and Sec.~\ref{sec:ja7} formalize task representations as action preconditions and contribution dependencies, and cross-training studies have verified mental-model alignment and post-training collaboration improvement \cite{NikolaidisShah2013CrossTraining}. This evidence has not yet separated changes in partner behavior from changes in task dependencies, leaving it unclear whether improvement stems from familiarity with action sequences or from understanding which conditions license the next step. Within a familiar partner and workflow, both forms of learning can yield comparable performance, but shift changes, task reorganization, or equipment changes require different knowledge to be retained and revised. If evaluation always centers on action precision and completion speed within the trained workflow, it cannot determine whether the robot can organize tasks under new conditions or whether the human can understand and support a new arrangement. Future work can construct transfer conditions that keep the basic actions unchanged but rearrange which actions depend on which---drawing on the test logic of holding primitives constant while varying their combinations \cite{LakeBaroni2018Systematicity}. With matched training volume, crossing ``task relationships unchanged but partner behavior changed'' with ``actions still familiar but task preconditions changed'' and comparing the effects of transferring operation history versus transferring dependency knowledge will isolate what transfers. Separate measures of the robot's precondition judgments and replanning, the human's action prediction and error correction, and the team's re-coordination cost will identify which knowledge transfers and which must be re-established.

\subsection{Pause, Handoff, and Withdrawal}
Pausing, handoff, and withdrawal must become collaboration options with explicit task state, responsibility attribution, and resumption conditions. Studies in Sec.~\ref{sec:ja2-commitment} discuss cooperation maintenance primarily through post-failure feedback, sustained commitment, and workaround behavior, while field studies reveal the difficulty of transferring knowledge and responsibility when personnel change \cite{index2,index14,index15}. These two bodies of evidence have not been connected under comparable conditions to test whether, when a human actually chooses to pause, adjust the division of labor, or withdraw, the robot can properly transfer unfinished work and resume cooperation once both parties are ready again. Pending items, unmet preconditions, and unconfirmed decisions all require someone to remain responsible; unclear handoff can cause omissions, duplicate execution, or repeated help requests directed at a person who has already exited. Duration of participation is simultaneously affected by task pressure and unfinished commitments, so meaningful sustained cooperation requires enabling responsibility transfer and letting the person autonomously decide whether to re-engage afterward. Future work can draw on structured clinical handoff protocols that organize current state, pending items, contingency plans, and confirmation by the receiver \cite{Starmer2014Handoff}, and translate this structure into a handoff record the robot can present and the incoming operator can confirm. Providing all conditions with the same pause and withdrawal options, the study can compare whether explicitly confirming the receiver, pending items, and resumption conditions---and maintaining a clear paused state when no receiver is available---reduces post-handoff omissions and duplicate work. Evaluation should separately record omissions and duplicate work after handoff, residual human burden, whether resumption aligns with both parties' intentions, and actual subsequent re-engagement choices.

\subsection{Asymmetric Dependence Among HRC Subcodes}

``Unidirectional dependence'' means that when one subcode appears, another usually appears as well, but not necessarily the other way around. For example, $P(Division | Deictic Cues)=100\%$ means that every paper containing Deictic Cues also contains Division. This indicates an asymmetric association, not a causal relationship.
The three subcodes most often depended on are 7.1 Division, 6.3 Monitoring, and 6.2 Prediction. Together, they capture three basic questions in HRC: who does what, whether the collaboration is still proceeding as expected, and what is likely to happen next.

7.1 Division is the main depended-on subcode because many mechanisms become meaningful only within a concrete contribution structure. Deictic Cues, Spatial Anchoring, Confirmation, Clarification, Communication, and Sustained Commitment often specify who should act, who should receive an action, or when a task should be handed over. For example, the robot navigates and points to the target while the human performs the physical operation~\cite{index9}; the robot delivers blocks and the participant stacks them, while feedback helps maintain this division of work~\cite{index2}. Thus, Division can be a necessary condition for these mechanisms in particular system designs, although Division itself does not require them.

6.3 Monitoring is depended on by mechanisms that must first detect a deviation before responding. Retry, Replanning, Assistance, Clarification, and repair-oriented Communication all require information about whether an action, plan, or interaction is still on track. In the surgical studies, the Mako Product Specialist monitors boundaries, sequence, and alignment, and this monitoring supports diagnosis and reset when tracking or registration fails~\cite{index14, index15}. Monitoring therefore functions as the informational basis for repair and intervention.

6.2 Prediction connects information about the partner to subsequent action. Partner, Language, Sharing, Confirmation, Control, and Responsibility can provide evidence for anticipating a partner’s action, preference, intention, or possible takeover. For example, the robot uses the human’s beliefs and desires to predict preferences and subsequent choices~\cite{index51}.

In addition to these three depended-on subcodes, some subcodes show extremely high levels of dependence on other subcodes. Most notably, Deictic Cues, Sustained Commitment, and Clarification depend on Division in 100\% of their co-occurring cases; Communication and Confirmation depend on Division in 94\% of cases. These relationships suggest that, in the current literature, Deictic Cues, Sustained Commitment, and Clarification are usually not isolated communicative phenomena. Instead, they tend to occur within an existing structure of role division and task handover.

\subsection{Common Execution Architecture for HRC}
JA6 Partner-Action Processing, JA7 Task Sharing and Complementary Contribution, and JA8 Action Coordination and Coordination Smoothers appear in 79, 77, and 73 papers respectively. It can be interpreted as the most characteristic ``online execution triad'' in HRC. Together, they address an engineering-friendly control problem: JA6a action observation and JA6b action prediction reduce uncertainty about what a partner is doing or will do; JA7, especially Division, clarifies who does what; and JA8 connects the two actions through temporal, spatial, sequential, and sensorimotor coordination, including force coupling. This explains their recurrence in handovers, collaborative transport, assembly, navigation, and co-creative tasks, particularly in short, task-oriented experiments.

This concentration also reveals a default assumption in the literature: once partner-action processing, task sharing, and smooth action coordination are in place, cooperation is often treated as established. Questions of continued willingness to cooperate, decision authority to revise a plan, and responsibility for breakdowns remain outside the execution architecture. Future work should connect this chain to Common Ground and Grounding, Breakdown Detection and Repair, and the Sense of Individual and Joint Agency, including joint-outcome ownership.

%% file: sections/8_limitations.tex
\section{Limitations}


Several limitations arise from our literature collection and assessment process. First, HRC spans multiple disciplines, and our focus on selected databases and venues may have underrepresented relevant work from fields such as psychology, sociology, and communication, especially when different terminology was used. This may affect our assessment of which collaborative mechanisms are well studied or understudied. Future reviews could broaden interdisciplinary coverage to test the robustness of these findings.



Second, studies near the inclusion--exclusion boundary and constructs expressed only implicitly required researcher interpretation. This was particularly relevant for shared intention and cooperative commitment (JA2), which were often inferred from interaction behavior rather than directly measured. Future work could distinguish explicit from inferred evidence and conduct sensitivity analyses to assess how such coding decisions affect the conclusions.



Finally, our theoretical framing also limits what can be captured. Joint Action theory provides a useful structure for organizing HRC research, but phenomena that do not map clearly onto existing constructs may be overlooked. Future studies could use more inductive analyses to identify such phenomena and assess whether additional constructs or complementary theoretical perspectives are needed.


%% file: sections/9_conclusion.tex
\section{Conclusion}

This review examined how Joint Action constructs are instantiated in empirical human--robot collaboration research. Across 83 papers, we identified nine constructs and 29 subcodes spanning Representation, Prediction, and Integration, and analyzed their prevalence and co-occurrence. The literature emphasizes partner-action prediction, task division, and coordination, while shared intention, commitment, and some representational processes receive less attention. These findings highlight gaps between successful task execution and stronger evidence of shared understanding and agency. Our framework provides a structured basis for comparing HRC systems and identifying underexplored directions for future research.

%% file: sections/z_appendix.tex
\section{Retrieval Databases, Queries and Screening Results}
\begin{table*}[h]
\centering
\caption{
Overview of databases, search queries, and screening results.
The temporal scope of the search covers literature published between 2021 and 2026.
No duplicates were identified after merging the retrieved records.
}
\label{tab:retrieval}
\footnotesize
\setlength{\tabcolsep}{4pt}
\renewcommand{\arraystretch}{1.10}

\begin{tabularx}{\textwidth}{
    >{\raggedright\arraybackslash}p{0.18\textwidth}
    >{\raggedright\arraybackslash}X
    >{\centering\arraybackslash}p{0.08\textwidth}
    >{\centering\arraybackslash}p{0.08\textwidth}
}
\toprule
\textbf{Database} &
\textbf{Search Query} &
\textbf{Retrieved} &
\textbf{Included} \\
\midrule

ACM DL &
$(robot^{*} \ \mathrm{AND}\ human)$
$\mathrm{AND}$
$(collaborat^{*} \ \mathrm{OR}\ cooperat^{*} \ \mathrm{OR}\ communicat^{*} \ \mathrm{OR}\ coordinat^{*})$,
searched in title, abstract, and keywords. &
870 &
42 \\

\addlinespace

RO-MAN (IEEE Xplore) &
Equivalent query as above, with wildcard terms expanded into corresponding word forms and searched in document title, abstract, and author keywords. &
378 &
27 \\

\addlinespace

International Journal of Social Robotics (Scopus) &
\texttt{TITLE-ABS-KEY}
using the same conceptual query as above. &
201 &
11 \\

Science Robotics (Scopus) &
Same query as above. &
51 &
1 \\

Computers in Human Behavior (Scopus) &
Same query as above. &
21 &
2 \\

Nature (Scopus) &
Same query as above. &
2 &
0 \\

\midrule
\textbf{Total} &
&
\textbf{1523} &
\textbf{83} \\

\bottomrule
\end{tabularx}
\end{table*}

\section{Pilot Sampling and Coding Framework}
\label{app:pilot-coding}

This appendix reports the pilot sampling design, the development and held-out samples, and the operational coding rules used in the analysis.


\subsection{Pilot Sampling Design}
\label{app:pilot-sampling}

Table~\ref{tab:app_sampling_matrix} lists the six sampling dimensions used to construct the pilot, Table~\ref{tab:app_boundary_pressure} details the boundary-pressure pairs, Table~\ref{tab:app_sampling_design} summarizes the three development rounds and the held-out stage, and Tables~\ref{tab:diagnostic_papers} and~\ref{tab:app_heldout_sample} report the resulting development and held-out samples.

\begin{table*}[t]
\centering
\caption{Pilot sampling dimensions and purposes.}
\label{tab:app_sampling_matrix}
\footnotesize
\setlength{\tabcolsep}{4pt}
\renewcommand{\arraystretch}{1.10}

\begin{tabularx}{\textwidth}{
    >{\raggedright\arraybackslash}p{3.4cm}
    >{\raggedright\arraybackslash}p{7.0cm}
    >{\raggedright\arraybackslash}X
}
\toprule
\textbf{Sampling Dimension}
& \textbf{Cues or Levels}
& \textbf{Purpose} \\
\midrule

Author terminology
& commitment/plan; attention/gaze/reference; common ground; prediction; task allocation; coordination; failure/recovery; agency
& Ensure coverage of different conceptual vocabularies. \\

Apparent coordination structure
& planned-like; emergent-like; hybrid-like; unclear
& Avoid over-representing a single coordination form. \\

Task form
& planning/allocation; assembly/handover; transport/co-manipulation; open-ended collaboration; failure recovery
& Cover different forms of task interdependence. \\

Interaction channel
& language; gaze/gesture; visual/AR; motion; haptic
& Cover different interaction modalities. \\

Evidence type
& system contribution; controlled user study; deployment or longer-term interaction
& Cover different study designs. \\

Boundary pressure
& See Table~\ref{tab:app_boundary_pressure}
& Include diagnostically difficult cases for testing coding boundaries. \\

\bottomrule
\end{tabularx}
\end{table*}

\begin{table*}[t]
\centering
\caption{Boundary-pressure cases used for pilot selection.}
\label{tab:app_boundary_pressure}
\footnotesize
\setlength{\tabcolsep}{4pt}
\renewcommand{\arraystretch}{1.10}

\begin{tabularx}{\textwidth}{
    >{\raggedright\arraybackslash}p{3.4cm}
    >{\raggedright\arraybackslash}p{6.0cm}
    >{\raggedright\arraybackslash}X
}
\toprule
\textbf{Boundary Pair}
& \textbf{Why Ambiguous}
& \textbf{Pilot-Selection Trigger} \\
\midrule

Intent expression/inference vs.\ shared intention
& Recognizing, predicting, or communicating intent does not necessarily establish shared intention.
& Include studies involving intent recognition, intent communication, or joint planning. \\

Gaze vs.\ joint attention
& Gaze may convey intent or social information without establishing a shared referent.
& Include gaze-, pointing-, and reference-related studies. \\

Information display vs.\ common ground
& Shared access to information does not necessarily establish common ground.
& Include shared-display, AR-guidance, and shared-awareness studies. \\

Task co-representation vs.\ partner prediction
& Representing the partner's task differs from predicting the partner's next action.
& Include studies using task or user models for prediction or planning. \\

Prediction vs.\ monitoring
& Prediction concerns future behavior, whereas monitoring concerns unfolding behavior or outcomes.
& Include studies involving prediction, online adaptation, or mismatch detection. \\

Task sharing vs.\ action coordination
& Task sharing concerns who contributes what; action coordination concerns how actions fit together during execution.
& Include allocation, lead/follow, handover, and co-manipulation studies. \\

Failure explanation vs.\ repair
& Identifying or explaining a failure does not necessarily restore the joint activity.
& Include failure-explanation, correction, recovery, and resumption studies. \\

Control allocation vs.\ sense of agency
& Objective control does not necessarily reflect an experienced sense of agency.
& Include mixed-initiative, leadership, and agency-measurement studies. \\

\bottomrule
\end{tabularx}
\end{table*}

\begin{table*}[t]
\centering
\caption{Development and held-out sampling design.}
\label{tab:app_sampling_design}
\footnotesize
\setlength{\tabcolsep}{4pt}
\renewcommand{\arraystretch}{1.10}

\begin{tabularx}{\textwidth}{
    >{\raggedright\arraybackslash}p{3.4cm}
    >{\centering\arraybackslash}p{1.0cm}
    >{\raggedright\arraybackslash}p{5.3cm}
    >{\raggedright\arraybackslash}X
}
\toprule
\textbf{Stage}
& \textbf{N}
& \textbf{Selection Logic}
& \textbf{Diagnostic Purpose} \\
\midrule

Round 1: Anchor cases
& 8
& Clear positive cases covering the main candidate constructs.
& Establish minimum evidence thresholds and key construct distinctions. \\

Round 2: Construct-boundary cases
& 8
& Papers close to neighboring constructs or prone to proxy-based false positives.
& Refine inclusion, exclusion, and co-coding rules. \\

Round 3: Stress-test cases
& 8
& Conceptually difficult cases involving failure/recovery, multimodality, complex coordination, or agency.
& Test framework coverage in difficult cases. \\

Held-out codebook test
& 12
& Selected with the same variation matrix and locked before Round 1.
& Assess transferability to papers not used in codebook development. \\

\bottomrule
\end{tabularx}

\end{table*}

\begin{table*}[t]
\centering
\caption{Development sample and diagnostic roles under the final codebook.}
\label{tab:diagnostic_papers}

\footnotesize
\setlength{\tabcolsep}{4pt}
\renewcommand{\arraystretch}{1.10}

\begin{tabularx}{\textwidth}{
    >{\centering\arraybackslash}p{1.8cm}
    >{\centering\arraybackslash}p{0.8cm}
    >{\raggedright\arraybackslash}p{4.3cm}
    >{\raggedright\arraybackslash}X
}
\toprule
\textbf{Round}
& \textbf{Year}
& \textbf{Author}
& \textbf{Diagnostic Purpose} \\
\midrule

\multirow{8}{1.8cm}{\centering\textbf{R1 Anchor}}
& 2021
& Vignolo et al.~\cite{index129}
& JA2 anchor; commitment vs.\ generic adaptation. \\

& 2025
& Chan et al.~\cite{index41}
& JA4/JA5 anchor; shared reference and grounding. \\

& 2025
& Sagheb et al.~\cite{index53}
& JA3 anchor; task/state representation vs.\ online processing. \\

& 2022
& Laplaza Galindo et al.~\cite{index106}
& JA6 anchor; future-action prediction. \\

& 2024
& Tassi et al.~\cite{index64}
& JA7 anchor; role organization vs.\ action coordination. \\

& 2025
& Allen et al.~\cite{index36}
& JA8 anchor; haptic and trajectory coordination. \\

& 2023
& Khanna et al.~\cite{index88}
& JA9 anchor; repair vs.\ explanation alone. \\

& 2024
& Yashin et al.~\cite{index80}
& JA10 anchor; agency vs.\ objective control. \\

\midrule

\multirow{8}{1.8cm}{\centering\textbf{R2 Boundary}}
& 2025
& Zhang et al.~\cite{index26}
& Indirect-speech boundary; shared intention and grounding. \\

& 2025
& Yang and Mavrogiannis~\cite{index40}
& Implicit signaling; responsiveness vs.\ action coordination. \\

& 2023
& Domínguez-Vidal and Sanfeliu~\cite{index92}
& Prediction vs.\ explicit signaling. \\

& 2024
& Domínguez-Vidal and Sanfeliu~\cite{index62}
& Anticipation vs.\ grounding and handover coordination. \\

& 2022
& Norman et al.~\cite{index111}
& Misunderstanding and correction; grounding/repair boundary. \\

& 2025
& Hellou et al.~\cite{index147}
& Misalignment correction; grounding/repair vs.\ trust outcomes. \\

& 2021
& Kimoto et al.~\cite{index114}
& Help solicitation; task/role representation and prediction. \\

& 2024
& Abe et al.~\cite{index68}
& Unanticipated action; monitoring, repair, and agency. \\

\midrule

\multirow{8}{1.8cm}{\centering\textbf{R3 Stress}}
& 2026
& Banerjee et al.~\cite{index1}
& Failure recovery; JA9 stress case. \\

& 2026
& Wang et al.~\cite{index5}
& Multimodal communication; grounding, coordination, and repair. \\

& 2026
& Vermeulen et al.~\cite{index14}
& Multi-actor collaboration stress case. \\

& 2026
& Cho et al.~\cite{index16}
& Emergent factory coordination stress case. \\

& 2026
& Rouchitsas et al.~\cite{index17}
& Spontaneous malfunction; repair vs.\ monitoring. \\

& 2025
& Kostolani et al.~\cite{index34}
& Multimodal failure recovery stress case. \\

& 2024
& Haripriyan et al.~\cite{index67}
& Broad team-dynamics stress case. \\

& 2024
& Deshpande et al.~\cite{index73}
& Open-ended co-creation stress case. \\

\bottomrule
\end{tabularx}
\end{table*}

\begin{table*}[h]
\centering
\caption{Held-out sample and the boundary or heterogeneity covered by each paper.}
\label{tab:app_heldout_sample}
\footnotesize
\setlength{\tabcolsep}{4pt}
\renewcommand{\arraystretch}{1.10}

\begin{tabularx}{\textwidth}{
    >{\centering\arraybackslash}p{0.06\textwidth}
    >{\raggedright\arraybackslash}p{0.24\textwidth}
    >{\raggedright\arraybackslash}X
}
\toprule
\textbf{Year}
& \textbf{Author}
& \textbf{Boundary or Heterogeneity Covered} \\
\midrule

2026
& Kim et al.~\cite{index2}
& Human-attributed failure; repair and agency. \\

2026
& Cai et al.~\cite{index9}
& Assistive navigation; shared reference and grounding. \\

2026
& Tulbure et al.~\cite{index12}
& Handover; task co-representation vs.\ partner-action processing. \\

2025
& Chang et al.~\cite{index18}
& Open-ended dance; emergent coordination and agency. \\

2025
& Pietras et al.~\cite{index27}
& Longer-term co-adaptation; grounding and role adjustment. \\

2025
& Bossema et al.~\cite{index45}
& Open-ended drawing; task representation and coordination. \\

2025
& Domínguez-Vidal and Sanfeliu~\cite{index58}
& Multimodal prediction vs.\ explicit signaling. \\

2024
& Ho et al.~\cite{index71}
& Multiparty learning; role representation and coordination. \\

2023
& Schreiter et al.~\cite{index83}
& Communication modality; shared reference and coordination. \\

2023
& Pelikan and Hofstetter~\cite{index95}
& Interaction delays; monitoring vs.\ timing coordination. \\

2023
& Al-Saadi et al.~\cite{index98}
& Physical conflict; coordination and repair. \\

2026
& Yu et al.~\cite{index140}
& Creative collaboration; grounding and agency. \\

\bottomrule
\end{tabularx}
\end{table*}

\subsection{Coding Manual and Boundary Rules}
\label{app:codebook}

Table~\ref{tab:app_detailed_codebook} presents the operational codebook for the nine primary Joint Action codes, and Table~\ref{tab:app_boundary_rules} lists the high-risk discriminant boundaries applied during full-text coding.

\begin{table*}[h]
\centering
\caption{Detailed theory-derived codebook for the nine primary Joint Action codes.}
\label{tab:app_detailed_codebook}

\footnotesize
\setlength{\tabcolsep}{4pt}
\renewcommand{\arraystretch}{1.10}

\begin{tabularx}{\textwidth}{
    >{\raggedright\arraybackslash}p{3.6cm}
    >{\raggedright\arraybackslash}X
    >{\raggedright\arraybackslash}X
    >{\raggedright\arraybackslash}X
}
\toprule
\textbf{Code}
& \textbf{Definition and Theory Source}
& \textbf{Include}
& \textbf{Exclude} \\
\midrule

\textbf{JA2 Shared Intention and Cooperative Commitment}
&
Joint planning and execution supported by shared intentional structure and continued cooperative commitment.~\cite{bratman1992,bratman2014}
&
Plan negotiation or alignment; mutual responsiveness; sustained commitment.
&
Intent recognition or display, simple task agreement, and task allocation alone are insufficient.
\\

\midrule

\textbf{JA3 Task Co-representation}
&
Representation of one's own task, the partner's task, joint-task rules, or relations among the partners' contributions.~\cite{sebanz2006,vesper2010,vesper2017}
&
Partner-task representation; joint-task models; interdependencies among subtasks.
&
Generic user models or prediction of the partner's next action alone are excluded. JA3 concerns task structure; JA6 concerns online partner-action processing.
\\

\midrule

\textbf{JA4 Joint Attention and Shared Reference}
&
Coordination of attention to establish a shared object, location, or event that supports subsequent joint action.~\cite{sebanz2006,vesper2017}
&
Gaze, pointing, or reference-resolution mechanisms that establish or confirm a shared referent.
&
Gaze estimation, intent inference, or social signaling without a shared referent is insufficient.
\\

\midrule

\textbf{JA5 Common Ground and Grounding}
&
Activity-relevant knowledge becomes mutually usable and is established or updated through evidence of understanding.~\cite{ClarkBrennan1991,Klein2005}
&
Grounding; acknowledgement or confirmation; representational alignment; updating.
&
One-way information presentation or coincidentally shared information without grounding evidence is insufficient.
\\

\midrule

\textbf{JA6 Partner-Action Processing}
&
Online perception, prediction, and monitoring of partner actions or joint outcomes in support of coordination.~\cite{sebanz2006,vesper2010,vesper2017}
&
Action observation; action prediction; action or outcome monitoring.
&
Generic perception or offline prediction disconnected from joint coordination is excluded. Prediction concerns future behavior; monitoring concerns unfolding behavior or outcomes.
\\

\midrule

\textbf{JA7 Task Sharing and Complementary Contribution}
&
Division of a joint task into interdependent roles or contributions that complement one another.~\cite{sebanz2006,vesper2017}
&
Task or role allocation; division of labor; lead/follow organization.
&
Independent parallel tasks and execution-level coordination alone are excluded. JA7 concerns ``who does what,'' whereas JA8 concerns how actions are coordinated.
\\

\midrule

\textbf{JA8 Action Coordination and Coordination Smoothers}
&
Coordination of actions during execution, including mechanisms that reduce coordination difficulty.~\cite{sebanz2006,vesper2010,vesper2017,knoblich2011}
&
Synchronization; handover timing; trajectory or haptic coupling; sensorimotor signaling; coordination smoothers.
&
Task allocation or performance outcomes without an analyzed coordination mechanism are insufficient.
\\

\midrule

\textbf{JA9 Breakdown Detection and Repair}
&
Detection of a breakdown in joint activity followed by a process that restores coordination.~\cite{ClarkBrennan1991,Klein2005}
&
Clarification or correction; re-grounding; replanning or re-coordination; observable resumption.
&
Error detection or explanation alone is insufficient when the joint activity is not restored.
\\

\midrule

\textbf{JA10 Sense of Individual and Joint Agency}
&
Experience or attribution of control, authorship, and ownership concerning one's contribution and the joint outcome.~\cite{sebanz2006,Pacherie2012,Loehr2022}
&
Direct agency ratings; intentional binding; reports of shared agency; joint-outcome ownership.
&
Autonomy level, objective control allocation, task performance, or trust alone are insufficient without direct agency evidence.
\\

\bottomrule
\end{tabularx}

\end{table*}

\begin{table*}[t]
\centering
\caption{High-risk discriminant boundaries used during full-text coding.}
\label{tab:app_boundary_rules}
\footnotesize
\setlength{\tabcolsep}{4pt}
\renewcommand{\arraystretch}{1.10}

\begin{tabularx}{\textwidth}{
    >{\raggedright\arraybackslash}p{4.0cm}
    >{\raggedright\arraybackslash}p{5.6cm}
    >{\raggedright\arraybackslash}X
}
\toprule
\textbf{Boundary}
& \textbf{High-Risk Error}
& \textbf{Operational Decision} \\
\midrule

Intent expression/inference vs.\ shared intention
& Intent recognition, prediction, or communication may be mistaken for shared intention.
& Code JA2 only with evidence of joint planning, mutual responsiveness, or cooperative commitment. \\

Gaze vs.\ joint attention
& Gaze may convey intent or social information without establishing a shared referent.
& Code JA4 only when gaze or gesture establishes or confirms a shared referent. \\

Information display vs.\ common ground
& Shared access to information may be mistaken for common ground.
& Code JA5 only when the information becomes mutually usable through grounding or evidence of understanding. \\

Task co-representation vs.\ partner prediction
& Representing the partner's task may be confused with predicting the partner's next action.
& Code task structure as JA3 and future-action prediction as JA6; co-code when both are supported. \\

Prediction vs.\ monitoring
& Anticipating future behavior may be confused with evaluating ongoing behavior.
& Keep prediction and monitoring as distinct processes within JA6. \\

Task sharing vs.\ action coordination
& Task division may be confused with execution-level coordination.
& JA7 concerns ``who does what''; JA8 concerns how actions are coordinated during execution. \\

Failure explanation vs.\ repair
& Detecting or explaining failure may be mistaken for repair.
& Code JA9 only when a recovery process restores the joint activity. \\

Control allocation vs.\ sense of agency
& Objective control may be mistaken for subjective agency.
& Code JA10 only when agency or authorship is directly measured or analyzed. \\

\bottomrule
\end{tabularx}
\end{table*}

\section{Details of the Survey Literature Statistics}

Table~\ref{tab:ja_all} lists the full reference set for each Joint Action dimension and subdimension discussed in the main text.

\begin{table*}[h]
  \caption{References of Joint Action Dimensions and Subdimensions.}
  \label{tab:ja_all}
\footnotesize
\setlength{\tabcolsep}{4pt}
\renewcommand{\arraystretch}{1.10}

  \begin{tabularx}{\textwidth}{
    p{0.22\textwidth}
    p{0.17\textwidth}
    p{0.055\textwidth}
    X
  }
    \toprule
    \textbf{Dimension} & \textbf{Subdimension} & \textbf{Number} & \textbf{Papers} \\
    \midrule

    \multirow{3}{0.22\textwidth}{Shared Intention and Cooperative Commitment}
    & Strategy Negotiation
    & 2
    & \cite{index24,index40} \\

    & Mutual Responsiveness
    & 3
    & \cite{index24,index129,index179} \\

    & Sustained Commitment
    & 7
    & \cite{index2,index14,index15,index67,index102,index109,index129} \\

    \midrule

    \multirow{4}{0.22\textwidth}{Task Co-representation}
    & Task
    & 15
    & \cite{index1,index9,index15,index19,index21,index27,index50,index73,index86,index104,index105,index122,index133,index146,index187} \\

    & Role
    & 4
    & \cite{index14,index47,index50,index61} \\

    & Partner
    & 10
    & \cite{index23,index31,index51,index53,index67,index71,index109,index111,index114,index148} \\

    & Action
    & 11
    & \cite{index12,index18,index33,index40,index45,index70,index80,index88,index117,index134,index140} \\

    \midrule

    \multirow{3}{0.22\textwidth}{Joint Attention and Shared Reference}
    & Language
    & 19
    & \cite{index1,index5,index21,index26,index31,index37,index39,index50,index51,index58,index62,index66,index92,index105,index109,index111,index114,index143,index148} \\

    & Deictic Cues
    & 10
    & \cite{index9,index21,index61,index67,index70,index83,index86,index88,index104,index147} \\

    & Spatial Anchoring
    & 12
    & \cite{index14,index15,index27,index37,index39,index41,index61,index71,index74,index83,index96,index128} \\

    \midrule

    \multirow{3}{0.22\textwidth}{Common Ground and Grounding}
    & Sharing
    & 6
    & \cite{index14,index26,index45,index51,index73,index96} \\

    & Confirmation
    & 17
    & \cite{index5,index9,index14,index26,index27,index41,index62,index104,index105,index109,index111,index129,index140,index143,index147,index156,index187} \\

    & Clarification
    & 6
    & \cite{index15,index21,index61,index86,index96,index109} \\

    \midrule

    \multirow{3}{0.22\textwidth}{Partner-Action Processing}
    & Perception
    & 48
    & \cite{index9,index12,index18,index19,index20,index21,index22,index23,index24,index30,index33,index34,index36,index40,index41,index45,index46,index47,index50,index58,index61,index62,index63,index64,index66,index68,index70,index73,index74,index80,index86,index88,index92,index96,index98,index105,index106,index119,index122,index123,index131,index134,index136,index137,index143,index156,index179,index187} \\

    & Anticipation
    & 54
    & \cite{index1,index5,index9,index12,index14,index15,index16,index17,index18,index20,index21,index23,index24,index26,index27,index28,index30,index31,index33,index36,index40,index41,index45,index47,index50,index51,index53,index58,index62,index63,index66,index67,index71,index73,index86,index92,index95,index98,index104,index106,index111,index114,index117,index122,index123,index128,index134,index140,index143,index146,index147,index148,index156,index179} \\

    & Monitoring
    & 51
    & \cite{index1,index9,index14,index15,index16,index17,index18,index23,index27,index28,index31,index33,index34,index37,index40,index41,index47,index50,index51,index61,index62,index63,index67,index68,index71,index73,index74,index86,index88,index92,index95,index96,index101,index102,index103,index104,index109,index111,index114,index117,index129,index131,index133,index140,index143,index146,index147,index148,index156,index179,index187} \\

    \midrule

    \multirow{3}{0.22\textwidth}{Task Sharing and Complementary Contribution}
    & Division
    & 62
    & \cite{index1,index2,index5,index9,index12,index14,index15,index17,index18,index19,index21,index22,index24,index26,index27,index28,index30,index31,index33,index34,index37,index41,index45,index47,index51,index53,index58,index61,index62,index64,index67,index68,index70,index71,index74,index80,index83,index86,index88,index92,index96,index101,index102,index103,index104,index109,index111,index119,index122,index128,index129,index133,index134,index137,index140,index143,index146,index147,index156,index179,index187,index188} \\

    & Leadership
    & 31
    & \cite{index9,index18,index20,index22,index24,index26,index36,index41,index46,index47,index50,index51,index58,index64,index67,index68,index71,index73,index92,index96,index98,index101,index105,index106,index109,index131,index136,index143,index156,index179,index188} \\

    & Assistance
    & 12
    & \cite{index16,index21,index31,index39,index63,index71,index103,index114,index131,index133,index148,index187} \\

    \midrule

    \multirow{4}{0.22\textwidth}{Action Coordination and Coordination Smoothers}
    & Signaling
    & 48
    & \cite{index2,index9,index14,index15,index19,index22,index26,index28,index30,index34,index36,index37,index39,index40,index41,index46,index47,index53,index58,index61,index63,index64,index66,index68,index73,index74,index83,index86,index88,index92,index95,index96,index101,index102,index104,index117,index119,index122,index123,index128,index129,index133,index134,index137,index140,index143,index147,index187} \\

    & Sequencing
    & 48
    & \cite{index2,index5,index9,index12,index16,index18,index19,index21,index23,index24,index26,index27,index28,index30,index33,index34,index37,index39,index41,index45,index46,index53,index58,index61,index67,index68,index71,index73,index74,index83,index86,index88,index95,index96,index101,index102,index103,index104,index105,index111,index114,index131,index133,index134,index137,index143,index147,index187} \\

    & Synchronization
    & 50
    & \cite{index2,index9,index12,index14,index15,index18,index21,index22,index24,index28,index30,index36,index37,index40,index41,index45,index46,index47,index50,index53,index58,index61,index62,index63,index64,index68,index73,index74,index80,index83,index86,index92,index98,index102,index103,index104,index105,index106,index117,index119,index122,index123,index128,index131,index133,index134,index136,index137,index140,index179} \\

    & Adaptation
    & 29
    & \cite{index5,index16,index20,index22,index23,index27,index28,index30,index34,index40,index46,index50,index58,index62,index67,index71,index80,index86,index88,index98,index104,index105,index114,index117,index119,index122,index129,index140,index179} \\

    \midrule

    \multirow{3}{0.22\textwidth}{Breakdown Detection and Repair}
    & Communication
    & 18
    & \cite{index1,index2,index5,index9,index17,index26,index31,index33,index51,index61,index63,index71,index96,index109,index111,index122,index147,index187} \\

    & Retry
    & 16
    & \cite{index2,index14,index15,index16,index27,index41,index66,index88,index95,index102,index104,index111,index129,index133,index143,index146} \\

    & Replanning
    & 22
    & \cite{index9,index16,index17,index20,index33,index34,index36,index41,index47,index50,index67,index68,index88,index96,index98,index109,index117,index131,index133,index143,index147,index179} \\

    \midrule

    \multirow{3}{0.22\textwidth}{Sense of Individual and Joint Agency}
    & Control
    & 18
    & \cite{index5,index17,index18,index20,index24,index47,index50,index58,index61,index67,index68,index71,index73,index92,index131,index134,index140,index143} \\

    & Attribution
    & 17
    & \cite{index5,index18,index24,index26,index45,index50,index61,index62,index66,index67,index70,index73,index80,index92,index105,index140,index188} \\

    & Responsibility
    & 11
    & \cite{index2,index16,index26,index47,index50,index58,index62,index66,index71,index92,index188} \\

    \bottomrule
  \end{tabularx}
\end{table*}

%% file: sample-base.bib
@String{Computing = "Computing" }

@String{Springer = "Springer-Verlag" }

@article{sebanz2006,
  title={Joint action: bodies and minds moving together},
  author={Sebanz, Natalie and Bekkering, Harold and Knoblich, G{\"u}nther},
  journal={Trends in cognitive sciences},
  volume={10},
  number={2},
  pages={70--76},
  year={2006},
  publisher={Elsevier}
}

@article{knoblich2011,
  title={Psychological research on joint action: theory and data},
  author={Knoblich, G{\"u}nther and Butterfill, Stephen and Sebanz, Natalie},
  journal={Psychology of learning and motivation},
  volume={54},
  pages={59--101},
  year={2011},
  publisher={Elsevier}
}

@article{vesper2010,
  title={A minimal architecture for joint action},
  author={Vesper, Cordula and Butterfill, Stephen and Knoblich, G{\"u}nther and Sebanz, Natalie},
  journal={Neural networks},
  volume={23},
  number={8-9},
  pages={998--1003},
  year={2010},
  publisher={Elsevier}
}

@article{vesper2017,
  title={Joint action: Mental representations, shared information and general mechanisms for coordinating with others},
  author={Vesper, Cordula and Abramova, Ekaterina and B{\"u}tepage, Judith and Ciardo, Francesca and Crossey, Benjamin and Effenberg, Alfred and Hristova, Dayana and Karlinsky, April and McEllin, Luke and Nijssen, Sari RR and others},
  journal={Frontiers in psychology},
  volume={7},
  pages={2039},
  year={2017},
  publisher={Frontiers Media SA}
}

@article{bratman1992,
  title={Shared cooperative activity},
  author={Bratman, Michael E},
  journal={The philosophical review},
  volume={101},
  number={2},
  pages={327--341},
  year={1992},
  publisher={JSTOR}
}

@book{bratman2014,
  title={Shared agency: A planning theory of acting together},
  author={Bratman, Michael E},
  year={2013},
  publisher={Oxford University Press}
}

@article{ClarkBrennan1991,
  title={Grounding in communication.},
  author={Clark, Herbert H and Brennan, Susan E},
  year={1991},
  publisher={American Psychological Association}
}

@article{Klein2005,
  title={Common ground and coordination in joint activity},
  author={Klein, Gary and Feltovich, Paul J and Bradshaw, Jeffrey M and Woods, David D},
  journal={Organizational simulation},
  volume={53},
  pages={139--184},
  year={2005}
}

@article{Pacherie2012,
  title={The Phenomenology of Joint Action: Self-Agency vs},
  author={Pacherie, Elisabeth},
  journal={Joint-Agency. Joint Attention: New Developments},
  year={2012}
}

@article{Loehr2022,
  title={The sense of agency in joint action: An integrative review},
  author={Loehr, Janeen D},
  journal={Psychonomic bulletin \& review},
  volume={29},
  number={4},
  pages={1089--1117},
  year={2022},
  publisher={Springer}
}

@article{HsiehShannon2005,
  title={Three approaches to qualitative content analysis},
  author={Hsieh, Hsiu-Fang and Shannon, Sarah E},
  journal={Qualitative health research},
  volume={15},
  number={9},
  pages={1277--1288},
  year={2005},
  publisher={Sage Publications Sage CA: Thousand Oaks, CA}
}

@article{Carroll2013,
  title={“Best fit” framework synthesis: refining the method},
  author={Carroll, Christopher and Booth, Andrew and Leaviss, Joanna and Rick, Jo},
  journal={BMC medical research methodology},
  volume={13},
  number={1},
  pages={37},
  year={2013},
  publisher={Springer}
}

@article{Gale2013,
  title={Using the framework method for the analysis of qualitative data in multi-disciplinary health research},
  author={Gale, Nicola K and Heath, Gemma and Cameron, Elaine and Rashid, Sabina and Redwood, Sabi},
  journal={BMC medical research methodology},
  volume={13},
  number={1},
  pages={117},
  year={2013},
  publisher={Springer}
}

@article{Suri2011,
  title={Purposeful sampling in qualitative research synthesis},
  author={Suri, Harsh},
  journal={Qualitative research journal},
  volume={11},
  number={2},
  pages={63--75},
  year={2011},
  publisher={Emerald Group Publishing Limited}
}

@article{Benoot2016,
  title={The use of purposeful sampling in a qualitative evidence synthesis: A worked example on sexual adjustment to a cancer trajectory},
  author={Benoot, Charlotte and Hannes, Karin and Bilsen, Johan},
  journal={BMC medical research methodology},
  volume={16},
  number={1},
  pages={21},
  year={2016},
  publisher={Springer}
}

@article{Villani2018,
  title={Survey on human--robot collaboration in industrial settings: Safety, intuitive interfaces and applications},
  author={Villani, Valeria and Pini, Fabio and Leali, Francesco and Secchi, Cristian},
  journal={Mechatronics},
  volume={55},
  pages={248--266},
  year={2018},
  publisher={Elsevier}
}

@article{Ajoudani2018,
  title={Progress and prospects of the human--robot collaboration},
  author={Ajoudani, Arash and Zanchettin, Andrea Maria and Ivaldi, Serena and Albu-Sch{\"a}ffer, Alin and Kosuge, Kazuhiro and Khatib, Oussama},
  journal={Autonomous robots},
  volume={42},
  number={5},
  pages={957--975},
  year={2018},
  publisher={Springer}
}

@article{Mukherjee2022,
  title={A survey of robot learning strategies for human-robot collaboration in industrial settings},
  author={Mukherjee, Debasmita and Gupta, Kashish and Chang, Li Hsin and Najjaran, Homayoun},
  journal={Robotics and Computer-Integrated Manufacturing},
  volume={73},
  pages={102231},
  year={2022},
  publisher={Elsevier}
}

@article{Singh2026,
  title={Human--Robot Teaming: A Comprehensive Survey on Collaboration, Communication, and Cognition},
  author={Singh, Saurav and Rantanen, Esa and Heard, Jamison},
  journal={ACM Transactions on Human-Robot Interaction},
  volume={15},
  number={2},
  pages={1--48},
  year={2026},
  publisher={ACM New York, NY}
}

@article{Salehzadeh2022,
  title={Purposeful communication in human--robot collaboration: A review of modern approaches in manufacturing},
  author={Salehzadeh, Roya and Gong, Jiaqi and Jalili, Nader},
  journal={IEEE Access},
  volume={10},
  pages={129344--129361},
  year={2022},
  publisher={IEEE}
}

@article{ZhangDoyle2023,
  title={Integrating intention-based systems in human-robot interaction: a scoping review of sensors, algorithms, and trust},
  author={Zhang, Yifei and Doyle, Thomas},
  journal={Frontiers in Robotics and AI},
  volume={10},
  pages={1233328},
  year={2023},
  publisher={Frontiers Media SA}
}

@article{Glawe2026,
  title={Human autonomy and sense of agency in human-robot interaction: A systematic literature review},
  author={Glawe, Felix and Schmeckel, Tim and Brauner, Philipp and Ziefle, Martina},
  journal={International Journal of Social Robotics},
  volume={18},
  number={5},
  pages={64},
  year={2026},
  publisher={Springer}
}

@article{Curioni2018,
  title={Joint action in humans: A model for human-robot interactions},
  author={Curioni, Arianna and Knoblich, Gunther and Sebanz, Natalie and Goswami, A and Vadakkepat, P},
  journal={Humanoid robotics: A reference},
  pages={2149--2167},
  year={2019},
  publisher={Springer Dordrecht, The Netherlands}
}

@article{IqbalRiek2019,
  title={Human-robot teaming: Approaches from joint action and dynamical systems},
  author={Iqbal, Tariq and Riek, Laurel D},
  journal={Humanoid robotics: A reference},
  pages={2293--2312},
  year={2019},
  publisher={Springer Dordrecht}
}

@article{Belhassein2022,
  title={Addressing joint action challenges in HRI: Insights from psychology and philosophy},
  author={Belhassein, Kathleen and Fern{\'a}ndez-Castro, V{\'\i}ctor and Mayima, Amandine and Clodic, Aur{\'e}lie and Pacherie, Elisabeth and Guidetti, Mich{\`e}le and Alami, Rachid and Cochet, H{\'e}l{\`e}ne},
  journal={Acta Psychologica},
  volume={222},
  pages={103476},
  year={2022},
  publisher={Elsevier}
}

@article{Navare2026,
  title={Joint action with human versus robot partners across psychological domains: a systematic review},
  author={Navare, Uma Prashant and Marc, Am{\'e}lie and Mostafaoui, Ghil{\`e}s and Belkaid, Marwen},
  journal={Acta Psychologica},
  volume={269},
  pages={107410},
  year={2026},
  publisher={Elsevier}
}

@article{hirano2026ball,
  title={From ball to rover: Transformable palm-sized rover SORA-Q for autonomous lunar exploration},
  author={Hirano, D and Inazawa, M and Sutoh, M and Nagata, M and Yoneda, Y and Watanabe, K and Sawada, H and Sakoda, G and Abe, S and Homma, S},
  journal={Science Robotics},
  volume={11},
  number={115},
  pages={eaec8039},
  year={2026},
  publisher={American Association for the Advancement of Science}
}

@article{onishi2026being,
  title={Being Hugged by a Social Robot during Tele-Counseling Reduces Anxiety in Individuals with Autism},
  author={Onishi, Yuya and Sumioka, Hidenobu and Kumazaki, Hirokazu and Shiomi, Masahiro},
  journal={International Journal of Social Robotics},
  volume={18},
  number={2},
  pages={20},
  year={2026},
  publisher={Springer}
}

@inproceedings{maccio2024kinesthetic,
  title={Kinesthetic teaching in robotics: a mixed reality approach},
  author={Macci{\`o}, Simone and Shaaban, Mohamad and Carf{\`\i}, Alessandro and Mastrogiovanni, Fulvio},
  booktitle={2024 33rd IEEE international conference on robot and human interactive communication (ROMAN)},
  pages={1610--1617},
  year={2024},
  organization={IEEE}
}

@inproceedings{hald2021human,
  title={Human-robot trust assessment using top-down visual tracking after robot task execution mistakes},
  author={Hald, Kasper and Rehm, Matthias and Moeslund, Thomas B},
  booktitle={2021 30th IEEE International Conference on Robot \& Human Interactive Communication (RO-MAN)},
  pages={892--898},
  year={2021},
  organization={IEEE}
}

@article{page2021prisma,
  title={The PRISMA 2020 statement: an updated guideline for reporting systematic reviews},
  author={Page, Matthew J and McKenzie, Joanne E and Bossuyt, Patrick M and Boutron, Isabelle and Hoffmann, Tammy C and Mulrow, Cynthia D and Shamseer, Larissa and Tetzlaff, Jennifer M and Akl, Elie A and Brennan, Sue E and others},
  journal={bmj},
  volume={372},
  year={2021},
  publisher={British Medical Journal Publishing Group}
}

@inproceedings{gilbert2025human,
  title={Human-Autonomy Collaboration for Escaping Local Minima},
  author={Gilbert, Alia and Kaur, Gurnoor and Mendez, Kevin and Xie, Yule and Robert, Lionel and Tilbury, Dawn},
  booktitle={2025 34th IEEE International Conference on Robot and Human Interactive Communication (RO-MAN)},
  pages={1374--1379},
  year={2025},
  organization={IEEE}
}

@inproceedings{khan2025errors,
  title={Errors matter! The Influence of Error Frequency and Timing on Fluency and Workload in a Virtual Human-Cobot Collaboration Environment},
  author={Khan, Hamd Mehfooz and Arevalo Arboleda, Stephanie and Raake, Alexander},
  booktitle={Proceedings of the 2025 31st ACM Symposium on Virtual Reality Software and Technology},
  pages={1--8},
  year={2025}
}

@article{laplaza2025enhancing,
  title={Enhancing robotic collaborative tasks through contextual human motion prediction and intention inference},
  author={Laplaza, Javier and Moreno, Francesc and Sanfeliu, Alberto},
  journal={International Journal of Social Robotics},
  volume={17},
  number={10},
  pages={2077--2096},
  year={2025},
  publisher={Springer}
}

@article{perez2024experimental,
  title={Experimental assessment of human--robot teaming for multi-step remote manipulation with expert operators},
  author={P{\'e}rez-D’Arpino, Claudia and Khurshid, Rebecca P and Shah, Julie A},
  journal={ACM Transactions on Human-Robot Interaction},
  volume={13},
  number={3},
  pages={1--26},
  year={2024},
  publisher={ACM New York, NY}
}

@article{ramachandruni2024uhtp,
  title={UHTP: a user-aware hierarchical task planning framework for communication-free, mutually-adaptive human-robot collaboration},
  author={Ramachandruni, Kartik and Kent, Cassandra and Chernova, Sonia},
  journal={ACM Transactions on Human-Robot Interaction},
  volume={13},
  number={3},
  pages={1--27},
  year={2024},
  publisher={ACM New York, NY}
}

@inproceedings{fuster2025negotiation,
  title={Negotiation of assignation plans in human-robot team task scheduling},
  author={Fuster-Pal{\`a}, Llum and Dalmasso, Marc and Aubach-Altes, Artur and Izquierdo-Badiola, Silvia and Sanfeliu, Alberto and Garrell, Ana{\'\i}s},
  booktitle={2025 34th IEEE International Conference on Robot and Human Interactive Communication (RO-MAN)},
  pages={1222--1228},
  year={2025},
  organization={IEEE}
}

@inproceedings{naik2021multi,
  title={Multi-modal proactive approaching of humans for human-robot cooperative tasks},
  author={Naik, Lakshadeep and Palinko, Oskar and Bodenhagen, Leon and Kr{\"u}ger, Norbert},
  booktitle={2021 30th IEEE International Conference on Robot \& Human Interactive Communication (RO-MAN)},
  pages={323--329},
  year={2021},
  organization={IEEE}
}

@inproceedings{lima2023assistive,
  title={Assistive force control in collaborative human-robot transportation},
  author={Lima, Bruno GC and Ferrentino, Enrico and Chiacchio, Pasquale and Vento, Mario},
  booktitle={2023 32nd IEEE International Conference on Robot and Human Interactive Communication (RO-MAN)},
  pages={2646--2651},
  year={2023},
  organization={IEEE}
}

@inproceedings{shah2023robot,
  title={Robot adaptation under operator cognitive fatigue using reinforcement learning},
  author={Shah, Jay K and Yadav, Aakash and Hopko, Sarah K and Mehta, Ranjana K and Pagilla, Prabhakar R},
  booktitle={2023 32nd IEEE International Conference on Robot and Human Interactive Communication (RO-MAN)},
  pages={1467--1474},
  year={2023},
  organization={IEEE}
}

@inproceedings{roy2023need,
  title={I need your help... or do i? maintaining situation awareness through performative autonomy},
  author={Roy, Sayanti and Smith, Trey and Coltin, Brian and Williams, Tom},
  booktitle={Proceedings of the 2023 ACM/IEEE international conference on human-robot interaction},
  pages={122--131},
  year={2023}
}

@article{moon2021design,
  title={Design of hesitation gestures for nonverbal human-robot negotiation of conflicts},
  author={Moon, Ajung and Hashmi, Maneezhay and Loos, HF Machiel Van Der and Croft, Elizabeth A and Billard, Aude},
  journal={ACM Transactions on Human-Robot Interaction (THRI)},
  volume={10},
  number={3},
  pages={1--25},
  year={2021},
  publisher={ACM New York, NY, USA}
}

@article{roche2022study,
  title={Study of kinesthetic negotiation ability in lightweight comanipulative decision-making tasks: Design and study of a virtual partner based on human-human interaction observation},
  author={Roche, Lucas and Saint-Bauzel, Ludovic},
  journal={ACM Transactions on Human-Robot Interaction (THRI)},
  volume={11},
  number={2},
  pages={1--23},
  year={2022},
  publisher={ACM New York, NY}
}

@article{chiou2022towards,
  title={Towards human--robot teaming: Tradeoffs of explanation-based communication strategies in a virtual search and rescue task},
  author={Chiou, Erin K and Demir, Mustafa and Buchanan, Verica and Corral, Christopher C and Endsley, Mica R and Lematta, Glenn J and Cooke, Nancy J and McNeese, Nathan J},
  journal={International Journal of Social Robotics},
  volume={14},
  number={5},
  pages={1117--1136},
  year={2022},
  publisher={Springer}
}

@inproceedings{index1,
  title={A Human-in-the-Loop Confidence-Aware Failure Recovery Framework for Modular Robot Policies},
  author={Banerjee, Rohan and Palempalli, Krishna and Yang, Bohan and Fang, Jiaying and Abdullah, Alif and Silver, Tom and Dean, Sarah and Bhattacharjee, Tapomayukh},
  booktitle={Proceedings of the 21st ACM/IEEE International Conference on Human-Robot Interaction},
  pages={346--355},
  year={2026}
}

@inproceedings{index2,
  title={Affective and Cognitive Feedback from a Robot for Human-attributed Failure Handling},
  author={Kim, Jihwan and Jung, Myeongul and Kang, Donghyun and Rhee, Taehyun James and Kim, Dooyong and Kim, Kwanguk},
  booktitle={Proceedings of the 2026 CHI Conference on Human Factors in Computing Systems},
  pages={1--16},
  year={2026}
}

@inproceedings{index5,
  title={Bidirectional Human-Robot Communication for Physical Human-Robot Interaction},
  author={Wang, Junxiang and Wang, Cindy and Soltani Zarrin, Rana and Erickson, Zackory},
  booktitle={Proceedings of the 21st ACM/IEEE International Conference on Human-Robot Interaction},
  pages={572--580},
  year={2026}
}

@inproceedings{index9,
  title={Navigation beyond wayfinding: Robots collaborating with visually impaired users for environmental interactions},
  author={Cai, Shaojun and Janaka, Nuwan and Ram, Ashwin and Shehan, Janidu and Wan, Yingjia and Hara, Kotaro and Hsu, David},
  booktitle={Proceedings of the 21st ACM/IEEE International Conference on Human-Robot Interaction},
  pages={1201--1210},
  year={2026}
}

@inproceedings{index12,
  title={Task-Oriented Robot-Human Handovers on Legged Manipulators},
  author={Tulbure, Andreea and Scheidemann, Carmen and Steiner, Elias and Hutter, Marco},
  booktitle={Proceedings of the 21st ACM/IEEE International Conference on Human-Robot Interaction},
  pages={376--384},
  year={2026}
}

@inproceedings{index14,
  title={The Choreography of Care: An Ethnographic Study of Human-Robot Collaboration in Makoplasty Surgeries},
  author={Vermeulen, Jasper T and Dwyer, James L and Burden, Alan and Caldwell, Glenda and Teixeira, M{\"u}ge and Guertler, Matthias and Crawford, Ross},
  booktitle={Proceedings of the 2026 CHI Conference on Human Factors in Computing Systems},
  pages={1--19},
  year={2026}
}

@inproceedings{index15,
  title={The Invisible Work of Robotic Surgery: How Specialists Support, Shoulder, and Sustain Human-Robot Collaboration},
  author={Vermeulen, Jasper T and Caldwell, Glenda Amayo and Teixeira, M{\"u}ge Belek Fialho and Burden, Alan G and Guertler, Matthias},
  booktitle={Proceedings of the 2026 Designing Interactive Systems Conference},
  pages={4269--4283},
  year={2026}
}

@inproceedings{index16,
  title={Thing Ethnography in a Factory: Exploring Emergent and Dynamic Relations of Cobots and Workers},
  author={Cho, Hyungjun and Seo, Jiyeon Amy and Sohn, Yongjae and Lee, Hee Rin},
  booktitle={Proceedings of the 2026 CHI Conference on Human Factors in Computing Systems},
  pages={1--21},
  year={2026}
}

@inproceedings{index17,
  title={" What do I do now?": Spontaneous Human Responses to Robot Effectiveness and Efficiency Malfunctions in Collaborative Robotics},
  author={Rouchitsas, Alexandros and Niu, Xuezhi and Castellano, Ginevra and G{\"u}rd{\"u}r Broo, Didem},
  booktitle={Proceedings of the 2026 CHI Conference on Human Factors in Computing Systems},
  pages={1--17},
  year={2026}
}

@article{index18,
  title={A Constructed Response: Designing and Choreographing Robot Arm Movements in Collaborative Dance Improvisation},
  author={Chang, Xiaoyu and Zhang, Fan and Fu, Kexue and Diana, Carla and Ju, Wendy and Lc, Ray},
  journal={Proceedings of the ACM on Human-Computer Interaction},
  volume={9},
  number={7},
  pages={1--31},
  year={2025},
  publisher={ACM New York, NY, USA}
}

@inproceedings{index19,
  title={A Grammar-Based Communication Model for Conveying Design Intent in Human-Robot Teams},
  author={Tokac, Iremnur},
  booktitle={2025 34th IEEE International Conference on Robot and Human Interactive Communication (RO-MAN)},
  pages={872--877},
  year={2025},
  organization={IEEE}
}

@article{index20,
  title={A machine learning approach to resolving conflicts in physical human--robot interaction},
  author={Dincer, Enes Ulas and Al-Saadi, Zaid and Hamad, Yahya M and Aydin, Yusuf and Kucukyilmaz, Ayse and Basdogan, Cagatay},
  journal={ACM Transactions on Human-Robot Interaction},
  volume={14},
  number={2},
  pages={1--29},
  year={2025},
  publisher={ACM New York, NY}
}

@inproceedings{index21,
  title={A Multi-Modal Interaction Framework for Efficient Human-Robot Collaborative shelf-picking},
  author={Pathak, Abhinav and Venkatesan, Kalaichelvi and Taha, Tarek and Muthusamy, Rajkumar},
  booktitle={2025 34th IEEE International Conference on Robot and Human Interactive Communication (RO-MAN)},
  pages={2162--2169},
  year={2025},
  organization={IEEE}
}

@article{index22,
  title={A non-parametric approach to exploring and quantifying the information flow in human--robot collaboration},
  author={Giardini Lahr, Gustavo Jose and Sirintuna, Doganay and Tassi, Francesco and Amor, Heni Ben and Ajoudani, Arash},
  journal={ACM Transactions on Human-Robot Interaction},
  volume={15},
  number={1},
  pages={1--35},
  year={2025},
  publisher={ACM New York, NY}
}

@inproceedings{index23,
  title={An adaptive emotion-aware strategy for human-agent negotiation: Insights from real-world human-robot experiments},
  author={Keskin, Mehmet Onur and {\c{C}}akan, Umut and Aydo{\u{g}}an, Reyhan},
  booktitle={Proceedings of the 25th ACM International Conference on Intelligent Virtual Agents},
  pages={1--9},
  year={2025}
}

@inproceedings{index24,
  title={Assessing the Creative Role of Cobots in Human--Robot Artistic Co-Creation},
  author={Zhong, Wanheng and Teearavarunyou, Sakol},
  booktitle={Proceedings of the 2025 5th International Conference on Control and Intelligent Robotics},
  pages={129--135},
  year={2025}
}

@inproceedings{index26,
  title={Can you pass that tool?: Implications of indirect speech in physical human-robot collaboration},
  author={Zhang, Yan and Ratnayake, Tharaka Sachintha and Sew, Cherie and Knibbe, Jarrod and Goncalves, Jorge and Johal, Wafa},
  booktitle={Proceedings of the 2025 CHI Conference on Human Factors in Computing Systems},
  pages={1--18},
  year={2025}
}

@inproceedings{index27,
  title={Co-Adaptation in Human-Robot Training Scenarios},
  author={Pietras, Emilia and Kiefer, Bernd and Hall, Stephanie and Dhanda, Mandeep and Zhao, Haoruo and Dhokia, Vimal and Borzone, Guglielmo and Kr{\"u}ger, Norbert and Bodenhagen, Leon},
  booktitle={2025 34th IEEE International Conference on Robot and Human Interactive Communication (RO-MAN)},
  pages={1720--1726},
  year={2025},
  organization={IEEE}
}

@inproceedings{index28,
  title={Do Re Mi Fa So Pass the Tool: Using Melodic Prediction to Improve Human-Robot Fluency},
  author={Rogel, Amit and Yang, Qiaoyu and Hayley, Jack and Weinberg, Gil},
  booktitle={2025 34th IEEE International Conference on Robot and Human Interactive Communication (RO-MAN)},
  pages={199--206},
  year={2025},
  organization={IEEE}
}

@article{index30,
  title={Enhancing object release fluency in robot to human handover using proprioceptive and exteroceptive information},
  author={Penzotti, Mattia and Controzzi, Marco},
  journal={International Journal of Social Robotics},
  volume={17},
  number={10},
  pages={2123--2132},
  year={2025},
  publisher={Springer}
}

@article{index31,
  title={Enhancing Robot Assistive Behaviour by Mentalising User Intent and Beliefs with Reinforcement Learning},
  author={Andriella, Antonio and Falcone, Giovanni and Rossi, Silvia},
  journal={International Journal of Social Robotics},
  volume={17},
  number={10},
  pages={2293--2309},
  year={2025},
  publisher={Springer}
}

@article{index33,
  title={Evaluating a decision-making architecture in human-robot collaboration experiments},
  author={Sobhani, Mehdi and Giuliani, Manuel and Smith, Jim and Pipe, Anthony and Peer, Angelika},
  journal={International Journal of Social Robotics},
  volume={17},
  number={11},
  pages={2579--2600},
  year={2025},
  publisher={Springer}
}

@inproceedings{index34,
  title={" Fluency in Failure": The Impact of Interaction Modalities on Human-Robot Collaboration in Error Recovery},
  author={Kostolani, David and Brenter, Bernd and Schlund, Sebastian},
  booktitle={2025 34th IEEE International Conference on Robot and Human Interactive Communication (RO-MAN)},
  pages={2529--2534},
  year={2025},
  organization={IEEE}
}

@inproceedings{index36,
  title={Haptic communication in human-human and human-robot co-manipulation},
  author={Allen, Katherine H and Rogers, Chris and Short, Elaine S},
  booktitle={2025 34th IEEE International Conference on Robot and Human Interactive Communication (RO-MAN)},
  pages={2535--2542},
  year={2025},
  organization={IEEE}
}

@inproceedings{index37,
  title={Help or hindrance: Understanding the impact of robot communication in action teams},
  author={Tanjim, Tauhid and George, Jonathan St and Ching, Kevin and Taylor, Angelique},
  booktitle={2025 34th IEEE International Conference on Robot and Human Interactive Communication (RO-MAN)},
  pages={1460--1465},
  year={2025},
  organization={IEEE}
}

@inproceedings{index39,
  title={Human-robot teaming field deployments: A comparison between verbal and non-verbal communication},
  author={Tanjim, Tauhid and Ekpo, Promise and Cao, Huajie and George, Jonathan St and Ching, Kevin and Lee, Hee Rin and Taylor, Angelique},
  booktitle={2025 34th IEEE International Conference on Robot and Human Interactive Communication (RO-MAN)},
  pages={1699--1704},
  year={2025},
  organization={IEEE}
}

@inproceedings{index40,
  title={Implicit communication in human-robot collaborative transport},
  author={Yang, Elvin and Mavrogiannis, Christoforos},
  booktitle={2025 20th ACM/IEEE International Conference on Human-Robot Interaction (HRI)},
  pages={23--33},
  year={2025},
  organization={IEEE}
}

@article{index41,
  title={Improving human--robot collaboration through augmented reality and eye gaze},
  author={Chan, Wesley P and Crouch, Morgan and Hoang, Khoa and Chen, Charlie and Robinson, Nicole and Croft, Elizabeth},
  journal={ACM Transactions on Human-Robot Interaction},
  volume={14},
  number={3},
  pages={1--19},
  year={2025},
  publisher={ACM New York, NY}
}

@inproceedings{index45,
  title={LLM-enhanced interactions in human-robot collaborative drawing with older adults},
  author={Bossema, Marianne and Allouch, Somaya Ben and Plaat, Aske and Saunders, Rob},
  booktitle={2025 34th IEEE International Conference on Robot and Human Interactive Communication (RO-MAN)},
  pages={700--707},
  year={2025},
  organization={IEEE}
}

@article{index46,
  title={Low-force physical human-robot interaction at the hands influence changes to gait coordination through sensorimotor engagement instead of direct gait propulsion},
  author={Wu, Mengnan and Ting, Lena H},
  journal={ACM Transactions on Human-Robot Interaction},
  volume={14},
  number={3},
  pages={1--21},
  year={2025},
  publisher={ACM New York, NY}
}

@inproceedings{index47,
  title={Meaningful Interactions in Human-AI Musicking},
  author={Miles, Oliver and Hazzard, Adrian and Moroz, Solomiya and Bishop, Laura and Vear, Craig},
  booktitle={Proceedings of the 20th International Audio Mostly Conference},
  pages={58--69},
  year={2025}
}

@article{index50,
  title={Perception--intention--action cycle in human--robot collaborative tasks: the collaborative lightweight object transportation use-case},
  author={Dom{\'\i}nguez-Vidal, Jose Enrique and Rodr{\'\i}guez, Nicol{\'a}s and Sanfeliu, Alberto},
  journal={International Journal of Social Robotics},
  volume={17},
  number={10},
  pages={1927--1956},
  year={2025},
  publisher={Springer}
}

@article{index51,
  title={Robot, did you read my mind? Modelling human mental states to facilitate transparency and mitigate false beliefs in human--robot collaboration},
  author={Angelopoulos, Georgios and Hellou, Mehdi and Vinanzi, Samuele and Rossi, Alessandra and Rossi, Silvia and Cangelosi, Angelo},
  journal={ACM Transactions on Human-Robot Interaction},
  volume={15},
  number={1},
  pages={1--29},
  year={2025},
  publisher={ACM New York, NY}
}

@article{index53,
  title={Should Collaborative Robots Be Transparent?},
  author={Sagheb, Shahabedin and Gandhi, Soham and Losey, Dylan P},
  journal={International Journal of Social Robotics},
  volume={17},
  number={6},
  pages={937--953},
  year={2025},
  publisher={Springer}
}

@article{index58,
  title={When the inference meets the explicitness or why multimodality can make us forget about the perfect predictor},
  author={Dom{\'\i}nguez-Vidal, Jose Enrique and Sanfeliu, Alberto},
  journal={International Journal of Social Robotics},
  volume={17},
  number={12},
  pages={2965--2980},
  year={2025},
  publisher={Springer}
}

@article{index61,
  title={Affinity Diagramming with a robot},
  author={Law, Matthew V and Nwagwu, Nnamdi and Kwatra, Amritansh and Lee, Seo-Young and Diangelis, Daniel M and Yu, Naifang and Gonzalez-Pumariega, Gonzalo and Rajesh, Amit and Hoffman, Guy},
  journal={ACM Transactions on Human-Robot Interaction},
  volume={13},
  number={1},
  pages={1--41},
  year={2024},
  publisher={ACM New York, NY}
}

@inproceedings{index62,
  title={Anticipation and proactivity. unraveling both concepts in human-robot interaction through a handover example},
  author={Dom{\'\i}nguez-Vidal, Jose Enrique and Sanfeliu, Alberto},
  booktitle={2024 33rd IEEE International Conference on Robot and Human Interactive Communication (ROMAN)},
  pages={957--962},
  year={2024},
  organization={IEEE}
}

@article{index63,
  title={Does the social robot Nao facilitate cooperation in high functioning children with ASD?},
  author={Kostrubiec, Viviane and Lajunta, Chlo{\'e} and Paubel, Pierre-Vincent and Kruck, Jeanne},
  journal={International Journal of Social Robotics},
  volume={16},
  number={2},
  pages={281--298},
  year={2024},
  publisher={Springer}
}

@inproceedings{index64,
  title={Evaluating leadership roles in human-robot interaction via highly dynamic collaborative tasks},
  author={Tassi, Francesco and Lahr, Gustavo JG and Sirintuna, Doganay and Ajoudani, Arash},
  booktitle={2024 33rd IEEE International Conference on Robot and Human Interactive Communication (ROMAN)},
  pages={963--969},
  year={2024},
  organization={IEEE}
}

@article{index66,
  title={Generating pattern-based conventions for predictable planning in human--robot collaboration},
  author={Lohrmann, Clare and Stull, Maria and Roncone, Alessandro and Hayes, Bradley},
  journal={ACM Transactions on Human-Robot Interaction},
  volume={13},
  number={4},
  pages={1--23},
  year={2024},
  publisher={ACM New York, NY}
}

@inproceedings{index67,
  title={Human-robot action teams: A behavioral analysis of team dynamics},
  author={Haripriyan, Arthi and Jamshad, Rabeya and Ramaraj, Preeti and Riek, Laurel D},
  booktitle={2024 33rd IEEE International conference on robot and human interactive communication (ROMAN)},
  pages={1443--1448},
  year={2024},
  organization={IEEE}
}

@article{index68,
  title={Human understanding and perception of unanticipated robot action in the context of physical interaction},
  author={Abe, Naoko and Hu, Yue and Benallegue, Mehdi and Yamanobe, Natsuki and Venture, Gentiane and Yoshida, Eiichi},
  journal={ACM Transactions on Human-Robot Interaction},
  volume={13},
  number={1},
  pages={1--26},
  year={2024},
  publisher={ACM New York, NY}
}

@article{index70,
  title={Investigation of joint action in go/no-go tasks: Development of a human-like eye robot and verification of action space},
  author={Hayashi, Kotaro},
  journal={International Journal of Social Robotics},
  volume={16},
  number={9},
  pages={1919--1932},
  year={2024},
  publisher={Springer}
}

@inproceedings{index71,
  title={" It's Not a Replacement:" Enabling Parent-Robot Collaboration to Support In-Home Learning Experiences of Young Children},
  author={Ho, Hui-Ru and Hubbard, Edward M and Mutlu, Bilge},
  booktitle={Proceedings of the 2024 CHI Conference on Human Factors in Computing Systems},
  pages={1--18},
  year={2024}
}

@inproceedings{index73,
  title={Perceptions of interaction dynamics in co-creative ai: A comparative study of interaction modalities in drawcto},
  author={Deshpande, Manoj and Park, Jisu and Pait, Supratim and Magerko, Brian},
  booktitle={Proceedings of the 16th Conference on Creativity \& Cognition},
  pages={102--116},
  year={2024}
}

@inproceedings{index74,
  title={Presentation of robot-intended handover position using vibrotactile interface during robot-to-human handover task},
  author={Mohammed Zaffir, Muhammad Akmal Bin and Wada, Takahiro},
  booktitle={Proceedings of the 2024 ACM/IEEE international conference on human-robot interaction},
  pages={492--500},
  year={2024}
}

@article{index80,
  title={When is Human--Robot Joint Agency Effective? The Case of Cooperative Reaction Games},
  author={Yashin, Artem S and Stolyarova, Anastasiya N and Melnichuk, Eugeni V and Arinkin, Nikita A and Zinina, Anna A and Zhao, Darisy G and Kotov, Artemiy A},
  journal={International Journal of Social Robotics},
  volume={16},
  number={3},
  pages={635--644},
  year={2024},
  publisher={Springer}
}

@inproceedings{index83,
  title={Advantages of multimodal versus verbal-only robot-to-human communication with an anthropomorphic robotic mock driver},
  author={Schreiter, Tim and Morillo-Mendez, Lucas and Chadalavada, Ravi T and Rudenko, Andrey and Billing, Erik and Magnusson, Martin and Arras, Kai O and Lilienthal, Achim J},
  booktitle={2023 32nd IEEE International Conference on Robot and Human Interactive Communication (RO-MAN)},
  pages={293--300},
  year={2023},
  organization={IEEE}
}

@inproceedings{index86,
  title={Crafting with a robot assistant: use social cues to inform adaptive handovers in human-robot collaboration},
  author={Tian, Leimin and He, Kerry and Xu, Shiyu and Cosgun, Akansel and Kulic, Dana},
  booktitle={Proceedings of the 2023 ACM/IEEE International Conference on Human-Robot Interaction},
  pages={252--260},
  year={2023}
}

@inproceedings{index88,
  title={Effects of explanation strategies to resolve failures in human-robot collaboration},
  author={Khanna, Parag and Yadollahi, Elmira and Bj{\"o}rkman, M{\aa}rten and Leite, Iolanda and Smith, Christian},
  booktitle={2023 32nd IEEE International Conference on Robot and Human Interactive Communication (RO-MAN)},
  pages={1829--1836},
  year={2023},
  organization={IEEE}
}

@inproceedings{index92,
  title={Inference VS. Explicitness. Do we really need the perfect predictor? The human-robot collaborative object transportation case},
  author={Dom{\'\i}nguez-Vidal, Jose Enrique and Sanfeliu, Alberto},
  booktitle={2023 32nd IEEE International Conference on Robot and Human Interactive Communication (RO-MAN)},
  pages={1866--1871},
  year={2023},
  organization={IEEE}
}

@article{index95,
  title={Managing delays in human-robot interaction},
  author={Pelikan, Hannah and Hofstetter, Emily},
  journal={ACM Transactions on Computer-Human Interaction},
  volume={30},
  number={4},
  pages={1--42},
  year={2023},
  publisher={ACM New York, NY}
}

@inproceedings{index96,
  title={Navigating to Success in Multi-Modal Human-Robot Collaboration: Analysis and Corpus Release},
  author={Lukin, Stephanie M and Pollard, Kimberly A and Bonial, Claire and Hudson, Taylor and Artstein, Ron and Voss, Clare and Traum, David},
  booktitle={2023 32nd IEEE International Conference on Robot and Human Interactive Communication (RO-MAN)},
  pages={1859--1865},
  year={2023},
  organization={IEEE}
}

@inproceedings{index98,
  title={Resolving conflicts during human-robot co-manipulation},
  author={Al-Saadi, Zaid and Hamad, Yahya M and Aydin, Yusuf and Kucukyilmaz, Ayse and Basdogan, Cagatay},
  booktitle={Proceedings of the 2023 ACM/IEEE international conference on human-robot interaction},
  pages={243--251},
  year={2023}
}

@inproceedings{index101,
  title={The effect of robot skill level and communication in rapid, proximate human-robot collaboration},
  author={Lee, Kin Man and Krishna, Arjun and Zaidi, Zulfiqar and Paleja, Rohan and Chen, Letian and Hedlund-Botti, Erin and Schrum, Mariah and Gombolay, Matthew},
  booktitle={Proceedings of the 2023 ACM/IEEE International Conference on Human-Robot Interaction},
  pages={261--270},
  year={2023}
}

@inproceedings{index102,
  title={Understanding human-robot teamwork in the wild: The difference between success and failure for mobile robots in hospitals},
  author={Eriksen, Kristina Tornbjerg and Bodenhagen, Leon},
  booktitle={2023 32nd IEEE International Conference on Robot and Human Interactive Communication (RO-MAN)},
  pages={277--284},
  year={2023},
  organization={IEEE}
}

@inproceedings{index103,
  title={Verbally soliciting human feedback in continuous human-robot collaboration: Effects of the framing and timing of reminders},
  author={Candon, Kate and Zhou, Helen and Gillet, Sarah and V{\'a}zquez, Marynel},
  booktitle={Proceedings of the 2023 ACM/IEEE International Conference on Human-Robot Interaction},
  pages={290--300},
  year={2023}
}

@inproceedings{index104,
  title={Visual and haptic cues for human-robot handover},
  author={Costanzo, Marco and Natale, Ciro and Selvaggio, Mario},
  booktitle={2023 32nd IEEE International Conference on Robot and Human Interactive Communication (RO-MAN)},
  pages={2677--2682},
  year={2023},
  organization={IEEE}
}

@inproceedings{index105,
  title={Worth the wait: Understanding how the benefits of performative autonomy depend on communication latency},
  author={Silva, Rafael Sousa and Lieng, Michelle and Muly, Emil and Williams, Tom},
  booktitle={2023 32nd IEEE International Conference on Robot and Human Interactive Communication (RO-MAN)},
  pages={126--133},
  year={2023},
  organization={IEEE}
}

@inproceedings{index106,
  title={Context and intention for 3d human motion prediction: experimentation and user study in handover tasks},
  author={Laplaza, Javier and Garrell, Ana{\'\i}s and Moreno-Noguer, Francesc and Sanfeliu, Alberto},
  booktitle={2022 31st IEEE international conference on robot and human interactive communication (RO-MAN)},
  pages={630--635},
  year={2022},
  organization={IEEE}
}

@inproceedings{index109,
  title={Do children adapt their perspective to a robot when they fail to complete a task?},
  author={Yadollahi, Elmira and Couto, Marta and Dillenbourg, Pierre and Paiva, Ana},
  booktitle={Proceedings of the 21st Annual ACM Interaction Design and Children Conference},
  pages={341--351},
  year={2022}
}

@inproceedings{index111,
  title={Efficacy of a ‘misconceiving’robot to improve computational thinking in a collaborative problem solving activity: a pilot study},
  author={Norman, Utku and Chin, Alexandra and Bruno, Barbara and Dillenbourg, Pierre},
  booktitle={2022 31st IEEE International Conference on Robot and Human Interactive Communication (RO-MAN)},
  pages={1413--1420},
  year={2022},
  organization={IEEE}
}

@article{index114,
  title={Help-Estimator: Robot requests for help from humans by estimating a person’s subjective time},
  author={Kimoto, Mitsuhiko and Yasumatsu, Yuuki and Imai, Michita},
  journal={International Journal of Social Robotics},
  volume={14},
  number={3},
  pages={617--630},
  year={2022},
  publisher={Springer}
}

@article{index117,
  title={Joint action with artificial agents: Human-likeness in behaviour and morphology affects sensorimotor signaling and social inclusion},
  author={Ciardo, Francesca and De Tommaso, Davide and Wykowska, Agnieszka},
  journal={Computers in Human Behavior},
  volume={132},
  pages={107237},
  year={2022},
  publisher={Elsevier}
}

@inproceedings{index119,
  title={Productive inconvenience: Facilitating posture variability by stimulating robot-to-human handovers},
  author={Zolotas, Mark and Luo, Rui and Bazzi, Salah and Saha, Dipanjan and Mabulu, Katiso and Kloeckl, Kristian and Pad{\i}r, Ta{\c{s}}k{\i}n},
  booktitle={2022 31st IEEE International Conference on Robot and Human Interactive Communication (RO-MAN)},
  pages={122--128},
  year={2022},
  organization={IEEE}
}

@inproceedings{index122,
  title={Study on the Perception of Implicit Indication When Collaborating with an Artificial Agent: Example with the Japanese Rice Cake “Mochi” Making},
  author={Guinot, Lena and Ando, Kozo and Iwata, Hiroyasu},
  booktitle={Proceedings of the 10th International Conference on Human-Agent Interaction},
  pages={140--147},
  year={2022}
}

@inproceedings{index123,
  title={Task-consistent signaling motions for improved understanding in human-robot interaction and workspace sharing},
  author={Camblor, Benjamin and Benhabib, Nassim and Daney, David and Padois, Vincent and Salotti, Jean-Marc},
  booktitle={2022 17th ACM/IEEE International Conference on Human-Robot Interaction (HRI)},
  pages={275--283},
  year={2022},
  organization={IEEE}
}

@article{index128,
  title={What Is Happening Behind The Wall? Towards a Better Understanding of a Hidden Robot's Intent By Multimodal Cues},
  author={Kassem, Khaled and Ungerb{\"o}ck, Tobias and Wintersberger, Philipp and Michahelles, Florian},
  journal={Proceedings of the ACM on Human-Computer Interaction},
  volume={6},
  number={MHCI},
  pages={1--19},
  year={2022},
  publisher={ACM New York, NY, USA}
}

@article{index129,
  title={A humanoid robot’s effortful adaptation boosts partners’ commitment to an interactive teaching task},
  author={Vignolo, Alessia and Powell, Henry and Rea, Francesco and Sciutti, Alessandra and Mcellin, Luke and Michael, John},
  journal={ACM Transactions on Human-Robot Interaction (THRI)},
  volume={11},
  number={1},
  pages={1--17},
  year={2021},
  publisher={ACM New York, NY}
}

@inproceedings{index131,
  title={Exploring the role of social robot behaviors in a creative activity},
  author={Hu, Yaxin and Feng, Lingjie and Mutlu, Bilge and Admoni, Henny},
  booktitle={Proceedings of the 2021 ACM designing interactive systems conference},
  pages={1380--1389},
  year={2021}
}

@inproceedings{index133,
  title={Investigating human-robot cooperation in a hospital environment: Scrutinising visions and actual realisation of mobile robots in service work},
  author={Tornbjerg, Kristina and Kanstrup, Anne Marie and Skov, Mikael B and Rehm, Matthias},
  booktitle={Proceedings of the 2021 ACM Designing Interactive Systems Conference},
  pages={381--391},
  year={2021}
}

@article{index134,
  title={Methods for expressing robot intent for human--robot collaboration in shared workspaces},
  author={Lemasurier, Gregory and Bejerano, Gal and Albanese, Victoria and Parrillo, Jenna and Yanco, Holly A and Amerson, Nicholas and Hetrick, Rebecca and Phillips, Elizabeth},
  journal={ACM Transactions on Human-Robot Interaction (THRI)},
  volume={10},
  number={4},
  pages={1--27},
  year={2021},
  publisher={ACM New York, NY, USA}
}

@article{index136,
  title={Study of coordination between patients with schizophrenia and socially assistive robot during physical activity},
  author={Aubin, Lise and Mostafaoui, Ghil{\`e}s and Amiel, Chlo{\'e} and Serr{\'e}, H{\'e}l{\`e}ne and Capdevielle, Delphine and de Menibus, Ma{\"e}lane Hellouin and Boich{\'e}, Julie and Schmidt, Richard and Raffard, St{\'e}phane and Marin, Ludovic},
  journal={International Journal of Social Robotics},
  volume={13},
  number={7},
  pages={1625--1640},
  year={2021},
  publisher={Springer}
}

@article{index137,
  title={The relevance of signal timing in human-robot collaborative manipulation},
  author={Cini, Francesca and Banfi, Tommaso and Ciuti, Gastone and Craighero, Laila and Controzzi, Marco},
  journal={Science Robotics},
  volume={6},
  number={58},
  pages={eabg1308},
  year={2021},
  publisher={American Association for the Advancement of Science}
}

@inproceedings{index140,
  title={From Tool to Partner: Expressive Behaviors as the Bridge to Human-Robot Creative Collaboration},
  author={Yu, Jie and Zhang, Xinyi and Luo, Yaqiong and Lc, Ray and Wu, Kaishun and Tong, Xin},
  booktitle={Proceedings of the 2026 CHI Conference on Human Factors in Computing Systems},
  pages={1--22},
  year={2026}
}

@inproceedings{index143,
  title={If It’s Safe, Don’t Ask: Decreasing Frustration through User Involvement for Risky Robot Behaviors},
  author={Leusmann, Jan and Sch{\"o}mbs, Sarah and Diedrich, Maximilian and M{\"u}ller, Florian},
  booktitle={Proceedings of the 21st ACM/IEEE International Conference on Human-Robot Interaction},
  pages={904--913},
  year={2026}
}

@inproceedings{index146,
  title={Alignment Strategies for Language-Model-Driven Robots in Human-Robot Collaboration: Effectiveness and Impact on Trust},
  author={Goubard, Cedric and Demiris, Yiannis},
  booktitle={2025 34th IEEE International Conference on Robot and Human Interactive Communication (RO-MAN)},
  pages={924--930},
  year={2025},
  organization={IEEE}
}

@inproceedings{index147,
  title={Can You Handle The Truth? The Effects of Robots Correcting Users’ Misalignment on Trust and Perceived Social Competence},
  author={Hellou, Mehdi and Angelopoulos, Georgios and Vinanzi, Samuele and Rossi, Alessandra and Rossi, Silvia and Cangelosi, Angelo},
  booktitle={2025 34th IEEE International Conference on Robot and Human Interactive Communication (RO-MAN)},
  pages={2410--2417},
  year={2025},
  organization={IEEE}
}

@inproceedings{index148,
  title={Comparing cognitive and affective theory of mind for an assistive robotics application},
  author={Raggioli, Luca and Cantiello, Antimo and Esposito, Raffaella and Rossi, Alessandra and Rossi, Silvia},
  booktitle={Proceedings of the 33rd ACM Conference on User Modeling, Adaptation and Personalization},
  pages={22--31},
  year={2025}
}

@inproceedings{index156,
  title={" Once Upon a Time…": an Adaptive Robotic Behavior for Engaging Cooperative Storytelling},
  author={Barbato, Mario and Raggioli, Luca and Rossi, Silvia},
  booktitle={2025 34th IEEE International Conference on Robot and Human Interactive Communication (RO-MAN)},
  pages={2367--2374},
  year={2025},
  organization={IEEE}
}

@inproceedings{index179,
  title={A microsociological approach to understanding the boundary between robot cooperativeness and uncooperativeness in human-robot collaboration},
  author={Abe, Naoko and Rye, David and Loke, Lian},
  booktitle={2022 31st IEEE International Conference on Robot and Human Interactive Communication (RO-MAN)},
  pages={1085--1092},
  year={2022},
  organization={IEEE}
}

@inproceedings{index187,
  title={Design and evaluation of a framework for reciprocal speech interaction in human-robot collaboration},
  author={Bolano, Gabriele and Iviani, Lawrence and Roennau, Arne and Dillmann, Ruediger},
  booktitle={2021 30th IEEE International Conference on Robot \& Human Interactive Communication (RO-MAN)},
  pages={806--812},
  year={2021},
  organization={IEEE}
}

@article{index188,
  title={Effect of relative status on responsibility attributions in human--robot collaboration: Mediating role of sense of responsibility and moderating role of power distance orientation},
  author={Lei, Xin and Rau, Pei-Luen Patrick},
  journal={Computers in Human Behavior},
  volume={122},
  pages={106820},
  year={2021},
  publisher={Elsevier}
}

@article{goodrich2008human,
  title={Human--robot interaction: a survey},
  author={Goodrich, Michael A and Schultz, Alan C},
  journal={Foundations and trends{\textregistered} in human--computer interaction},
  volume={1},
  number={3},
  pages={203--275},
  year={2008},
  publisher={Now Publishers}
}

@article{li2015benefit,
  title={The benefit of being physically present: A survey of experimental works comparing copresent robots, telepresent robots and virtual agents},
  author={Li, Jamy},
  journal={International Journal of Human-Computer Studies},
  volume={77},
  pages={23--37},
  year={2015},
  publisher={Elsevier}
}

@article{riek2012wizard,
  title={Wizard of oz studies in hri: a systematic review and new reporting guidelines},
  author={Riek, Laurel D},
  journal={Journal of human-robot interaction},
  volume={1},
  number={1},
  pages={119--136},
  year={2012},
  publisher={Journal of Human-Robot Interaction Steering Committee}
}

@inproceedings{scholtz2003theory,
  title={Theory and evaluation of human robot interactions},
  author={Scholtz, Jean},
  booktitle={36th Annual Hawaii International Conference on System Sciences, 2003. Proceedings of the},
  pages={10--pp},
  year={2003},
  organization={IEEE}
}

@inproceedings{bradshaw2009tools,
  title={From tools to teammates: Joint activity in human-agent-robot teams},
  author={Bradshaw, Jeffrey M and Feltovich, Paul and Johnson, Matthew and Breedy, Maggie and Bunch, Larry and Eskridge, Tom and Jung, Hyuckchul and Lott, James and Uszok, Andrzej and van Diggelen, Jurriaan},
  booktitle={International conference on human centered design},
  pages={935--944},
  year={2009},
  organization={Springer}
}

@misc{iso8373,
  author       = {{International Organization for Standardization}},
  title        = {{ISO 8373:2021 Robotics --- Vocabulary}},
  year         = {2021},
  howpublished = {International Standard},
  url          = {https://www.iso.org/standard/75539.html}
}

@inproceedings{hoffman2004robots,
  title={Robots that Work in Collaboration with People.},
  author={Hoffman, Guy and Breazeal, Cynthia},
  booktitle={AAAI Technical Report (5)},
  pages={25--30},
  year={2004}
}

@article{bauer2008human,
  title={Human--robot collaboration: a survey},
  author={Bauer, Andrea and Wollherr, Dirk and Buss, Martin},
  journal={International Journal of Humanoid Robotics},
  volume={5},
  number={01},
  pages={47--66},
  year={2008},
  publisher={World Scientific}
}

@article{onnasch2021taxonomy,
  title={A taxonomy to structure and analyze human--robot interaction},
  author={Onnasch, Linda and Roesler, Eileen},
  journal={International Journal of Social Robotics},
  volume={13},
  number={4},
  pages={833--849},
  year={2021},
  publisher={Springer}
}

@article{Nikolaidis2017MutualAdaptation,
  title={Human-robot mutual adaptation in collaborative tasks: Models and experiments},
  author={Nikolaidis, Stefanos and Hsu, David and Srinivasa, Siddhartha},
  journal={The International Journal of Robotics Research},
  volume={36},
  number={5-7},
  pages={618--634},
  year={2017},
  publisher={SAGE Publications Sage UK: London, England}
}

@article{Morita2022Agency,
  title={Auditory feedback for enhanced sense of agency in shared control},
  author={Morita, Tomoya and Zhu, Yaonan and Aoyama, Tadayoshi and Takeuchi, Masaru and Yamamoto, Kento and Hasegawa, Yasuhisa},
  journal={Sensors},
  volume={22},
  number={24},
  pages={9779},
  year={2022},
  publisher={MDPI}
}

@article{Endsley1995SituationAwareness,
  title={Toward a theory of situation awareness in dynamic systems},
  author={Endsley, Mica R},
  journal={Human factors},
  volume={37},
  number={1},
  pages={32--64},
  year={1995},
  publisher={SAGE Publications Sage CA: Los Angeles, CA}
}

@article{TuckerEdmondson2003Learning,
  title={Why hospitals don't learn from failures: Organizational and psychological dynamics that inhibit system change},
  author={Tucker, Anita L and Edmondson, Amy C},
  journal={California management review},
  volume={45},
  number={2},
  pages={55--72},
  year={2003},
  publisher={SAGE Publications Sage CA: Los Angeles, CA}
}

@inproceedings{NikolaidisShah2013CrossTraining,
  title={Human-robot cross-training: Computational formulation, modeling and evaluation of a human team training strategy},
  author={Nikolaidis, Stefanos and Shah, Julie},
  booktitle={2013 8th ACM/IEEE international conference on human-robot interaction (HRI)},
  pages={33--40},
  year={2013},
  organization={IEEE}
}

@inproceedings{LakeBaroni2018Systematicity,
  title={Generalization without systematicity: On the compositional skills of sequence-to-sequence recurrent networks},
  author={Lake, Brenden and Baroni, Marco},
  booktitle={International conference on machine learning},
  pages={2873--2882},
  year={2018},
  organization={PMLR}
}

@article{Starmer2014Handoff,
  title={Changes in Medical Errors after Implementation of a Handoff Program},
  author={Starmer, Amy J. and Spector, Nancy D. and Srivastava, Rajendu and West, Daniel C. and Rosenbluth, Glenn and Allen, April D. and Noble, Elizabeth L. and Tse, Lisa L. and Dalal, Anuj K. and Keohane, Carol A. and Lipsitz, Stuart R. and Rothschild, Jeffrey M. and Wien, Matthew F. and Yoon, Catherine S. and Zigmont, Katherine R. and Wilson, Karen M. and O'Toole, Jennifer K. and Solan, Lauren G. and Aylor, Megan and Bismilla, Zia and Coffey, Maitreya and Mahant, Sanjay and Blankenburg, Rebecca L. and Destino, Lauren A. and Everhart, Jennifer L. and Patel, Shilpa J. and Bale, James F. and Spackman, Jaime B. and Stevenson, Adam T. and Calaman, Sharon and Cole, F. Sessions and Balmer, Dorene F. and Hepps, Jennifer H. and Lopreiato, Joseph O. and Yu, Clifton E. and Sectish, Theodore C. and Landrigan, Christopher P.},
  journal={New England Journal of Medicine},
  volume={371},
  number={19},
  pages={1803--1812},
  year={2014},
  publisher={Massachusetts Medical Society},
  doi={10.1056/NEJMsa1405556}
}
